\documentclass[pra, reprint]{revtex4-2}

\usepackage{placeins}
\usepackage{physics}
\usepackage{dsfont}

\usepackage[english]{babel}

\usepackage{amsmath}
\usepackage{graphicx}
\usepackage[colorlinks=true, allcolors=blue]{hyperref}
\usepackage{microtype}

\begin{document}
\author{Beatriz Darna}
\author{Marie Labeye}
\title{Effect of Physical Barriers on High-Harmonic Generation in Liquid-Phase}
\affiliation{CPCV, Département de chimie, École normale supérieure, PSL University, Sorbonne Université, CNRS, 75005 Paris, France}

\begin{abstract}
High-harmonic generation (HHG) in liquid-phase systems presents fundamental differences from gas and solid phases, notably a reduction in the harmonic cutoff, its apparent independence from laser parameters, and the appearance of a second plateau. In this work, we introduce a one-dimensional quantum model that incorporates physical boundaries to simulate the effect of solvation shells in liquid water by means of a complex absorbing potential (CAP). By systematically varying the position, width, and strength of the CAP, we explore how these physical constraints influence the HHG spectrum. Our findings confirm that spatial confinement, specifically the physical truncation of electron trajectories, is the primary factor governing the cutoff in liquid-phase HHG. This confinement leads to a cutoff that is independent of laser intensity and wavelength, aligning with experimental observations and existing hypotheses proposed by Mondal \textit{et al.} in 2023 and 2026. Furthermore, by implementing a semi-transparent CAP, we successfully reproduce the less intense second plateau observed in liquid-phase HHG. Time-frequency analysis confirms that this characteristic spectral feature is driven by partial wave-packet transmission through the solvation shells, which allows a surviving fraction of electrons to undergo longer continuum excursions. Our one-dimensional approach successfully isolates the fundamental role of physical boundaries in liquid-phase HHG, providing a clear and intuitive understanding of the underlying physics.
\end{abstract}

\maketitle

\section{INTRODUCTION}

High-harmonic generation (HHG) is a highly non-linear optical process in which intense laser-matter interaction produces a broad spectrum of frequencies, reaching up to the extreme ultraviolet (XUV) regime. First observed in gases during the 1980s \cite{journalarticle:McPherson.etal87,journalarticle:Ferray.etal88}, gas-phase HHG yields a characteristic spectrum featuring a broad non-perturbative ``plateau" of comparable harmonic intensities, ending in a sharp decay at the energy cutoff. This phenomenon is fundamentally understood through the semi-classical ``Three-Step Model'' (TSM) \cite{journalarticle:Corkum93} and its fully quantum formulation via the Strong-Field Approximation, which accounts for wave-packet spreading and quantum interference \cite{journalarticle:Lewenstein.etal94}. In this framework, HHG occurs via (i) tunnel ionization of an electron from its parent molecule, (ii) acceleration in the continuum by the driving laser field, and (iii) recombination with the parent ion to emit a high-energy photon. The TSM approximately predicts the cutoff energy and electron trajectory limits, both of which are strongly dependent on the amplitude and wavelength of the driving field. This theoretical framework has been successfully applied to multiple applications, ranging from attosecond pulse generation \cite{journalarticle:Paul.etal01,journalarticle:Hentschel.etal01a,journalarticle:Gaumnitz.etal17} and coherent XUV pulse shaping \cite{journalarticle:Wituschek.etal20,journalarticle:Yang.etal21} to ultrafast spectroscopy \cite{journalarticle:Corkum.Krausz07, journalarticle:Pertot.etal17,journalarticle:Yin.etal23}. HHG has also been discovered in the crystalline solid phase \cite{journalarticle:Ghimire.etal11}, opening a very active research field \cite{journalarticle:Luu.etal15,journalarticle:Guan.etal16,journalarticle:Ghimire.Reis19a,journalarticle:Goulielmakis.Brabec22}. In this case, HHG can also be described by an analogous three-step process in the momentum space, where the phenomenon is understood in terms of Bloch states \cite{journalarticle:Vampa.etal15a,journalarticle:Wu.etal16,journalarticle:Osika.etal17,journalarticle:Yue.Gaarde20}.

Despite the extensive theoretical and experimental progress achieved in both gaseous and solid media, the liquid phase remained largely unexplored due to severe technical and theoretical complexities. However, understanding this phase is fundamental, as the vast majority of biochemical processes, among others, occur in aqueous environments. The first HHG spectrum of liquid water, recorded in 2018 \cite{journalarticle:Luu.etal18}, opened the door to new strong-field investigations. This foundational study revealed that liquid-phase HHG is governed by underlying mechanisms fundamentally distinct from those in gases and solids. Subsequent efforts have aimed to better understand these processes \cite{journalarticle:Zeng.Bian20,journalarticle:Neufeld.etal22,journalarticle:Chen.Bian23,journalarticle:Ding.Bian26}. Notably, recent findings by Mondal \textit{et al.} \cite{journalarticle:Mondal.etal23, journalarticle:Mondal.etal26} demonstrated that the energy cutoff in liquid water is independent of both laser wavelength and intensity, a great contrast to the behavior of gases and solids. To explain this phenomenom, they proposed a modified TSM in which molecules in the liquid behave similarly to those in the gas phase, but with a strict spatial limit that ionized electrons cannot exceed. This spatial restriction justifies the observed intensity and wavelength independence. Furthermore, liquid water exhibits a very characteristic second plateau absent in the gas  and solid phase. These results strongly suggest that the solvation shells surrounding water molecules critically influence the HHG process, governing both the cutoff position and the emergence of this secondary plateau.

However, while Mondal's model relies on a purely semi-classical approach, the present work aims to demonstrate that physical barriers, specifically solvation shells, directly influence both the position and the number of cutoffs in the liquid-water HHG spectrum within a quantum framework. This study approaches the liquid environment by starting from an isolated gas-phase molecule and progressively introducing structural constraints through a one-dimensional (1D) quantum model. Despite its simplicity, a 1D approach is highly appropriate for this system, because the driving laser pulses are linearly polarized, the essential electron dynamics (tunneling, acceleration, and recombination) occur almost exclusively along the polarization axis. In addition, the low computational cost allows for the extensive calculations required to explore the underlying mechanisms of HHG. In these simulations, Complex Absorbing Potentials (CAPs) explicitly act as spatial limits for the electrons. By systematically varying their position, quantity, and strength, the model successfully mimics the solvation shells characteristic of a liquid. This framework enables the investigation of how these physical boundaries alter the emission spectrum, providing a precise analysis of their impact on the macroscopic HHG signal.

It is important to note that, given the one-dimensional nature of the employed code, this study does not aim to provide exact quantitative predictions, but rather to capture qualitatively the physical mechanism governing the system. The main goal is to isolate and study the fundamental influence of a physical barrier on the HHG process. Specifically, we aim to demonstrate a fundamental difference: while the laser drives the cutoff in gases, the physical barrier is the true limiting factor in confined systems. The remainder of this article is organized as follows. Section \ref{sec:computational_details} introduces the theoretical and computational framework. Within it, Section \ref{sec:1Dmodel} presents the 1D model and water potential, Section \ref{sec:efield} describes the driving laser pulses, Section \ref{sec:TDSE} explains how we numerically solve the Time-Dependent Schrödinger Equation (TDSE) and finally Sections \ref{sec:CAP} and \ref{sec:CAP2} detail the implementation and positioning of the CAPs. For the results and discussion, Section \ref{sec:water_molecule} establishes a reliable gas-phase baseline by analyzing HHG in an isolated water molecule. Following this, Section \ref{sec:1cap} examines the effect of a fully opaque physical barrier, specifically verifying the cutoff's independence from laser wavelength and intensity. Section \ref{sec:2caps} then introduces a semi-permeable barrier to demonstrate how partial wave-packet transmission explains the emergence of the second plateau observed in the liquid phase. Finally, the main results and conclusions are summarized in Section \ref{sec:conclusion}. Unless otherwise stated, atomic units are used throughout this article.

\section{COMPUTATIONAL DETAILS}
\label{sec:computational_details}

\subsection{One-Dimensional Model and Molecular Potential}
\label{sec:1Dmodel}

One-dimensional models are widely used in strong-field physics \cite{journalarticle:Guan.etal16,journalarticle:Yue.Gaarde20,journalarticle:Osika.etal17,thesis:Labeye18,journalarticle:Labeye.etal18}. Although they rely on several approximations to simplify the problem, their low computational cost is a great advantage. It allows for a vast number of calculations to be performed, enabling a deeper exploration of the underlying physical mechanisms. In addition, despite these simplifications, a 1D approach is highly relevant for this specific system. The driving laser pulses are linearly polarized, therefore the essential electron dynamics i.e. tunneling, acceleration, and recombination, occur almost exclusively along the polarization axis. Consequently, transverse effects are  negligible, making the 1D model not only computationally efficient but also physically adapted for the phenomena.

\begin{figure}
    \centering
    \includegraphics[width=\linewidth]{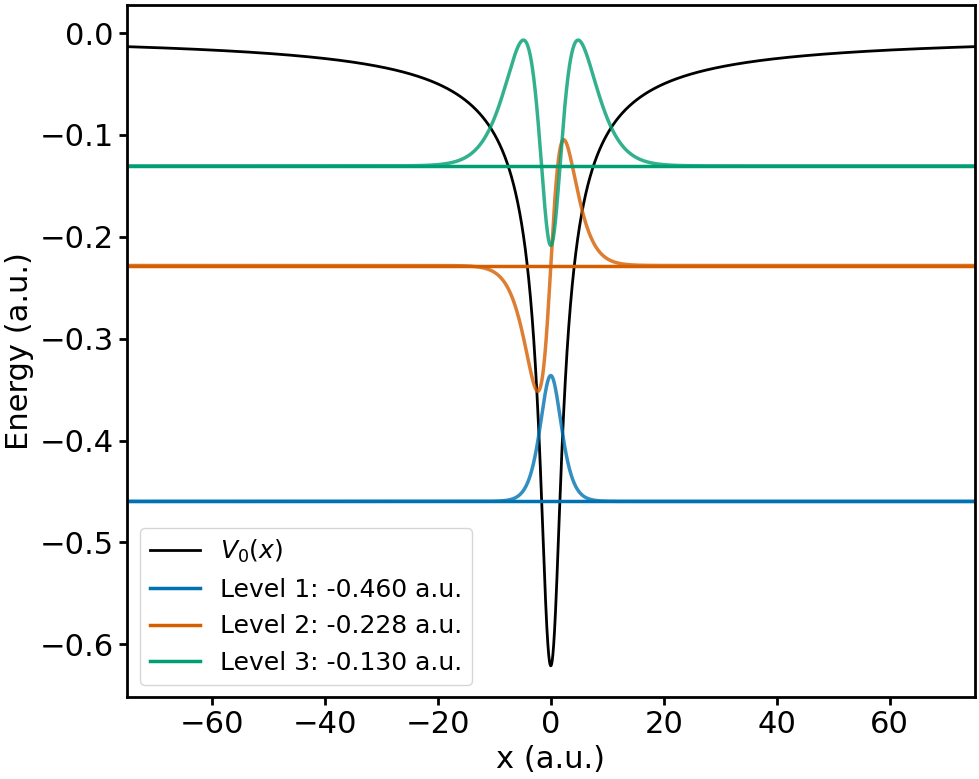}
    \caption{First three electronic bound states of the soft-Coulomb potential of a water molecule. The ground state energy corresponds to our ionisation potential ($I_\text{p}= 0.460$ a.u.).}
    \label{fig:potential}
\end{figure}

The model employed in this article is based on the Single Active Electron (SAE) approximation \cite{,journalarticle:Awasthi.etal08}. The active electron is initially trapped in a potential well, $V_0$, located at the center of the box, and it represents the water molecule. This potential is constructed from a combination of soft-Coulomb potentials. For the single isolated water molecule, the soft-Coulomb potential $V_0 (x)$ takes the following form:
\begin{multline}
V_0(x) = -\frac{Z_H}{\sqrt{{a_H}^2+(x - R/2)^2}} \\
-\frac{Z_O}{\sqrt{{a_O}^2+x^2}} \\
-\frac{Z_H}{\sqrt{{a_H}^2+(x + R/2)^2}}.
\end{multline}
where $Z_H$, $Z_O$, $a_H$ and $a_O$ correspond to the partial charge and regularization parameters of both the hydrogen and oxygen atoms, respectively, and $R$ is the distance between atoms of the same molecule in 1D.
The regularization parameters $a$ are chosen so that the ionization potential ($I_\text{p}$) matches the experimental value of the water molecule ($I_\text{p} =  12.6188 \pm (9 \times 10^{-4}) \text{ eV}$, or $\approx 0.4637$ a.u. \cite{journalarticle:Page.etal88}). Figure \ref{fig:potential} shows the potential $V_0$ for a single water molecule, along with its first three electronic states represented. The energy of the ground state (Level 1) closely matches this $I_\text{p}$.

\subsection{Electric Field}
\label{sec:efield}

\begin{figure}
    \centering
    \includegraphics[width=\linewidth]{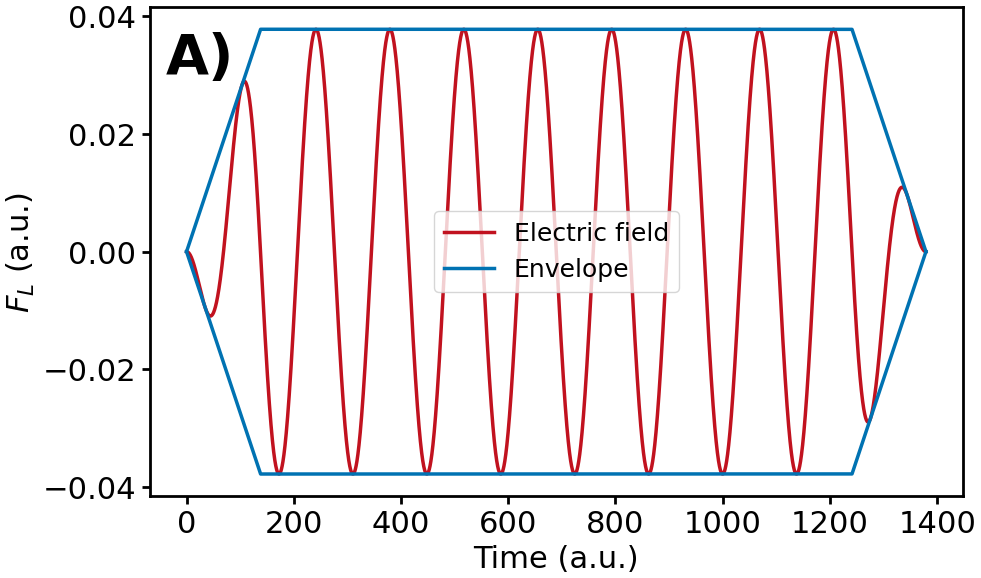}
    \includegraphics[width=\linewidth]{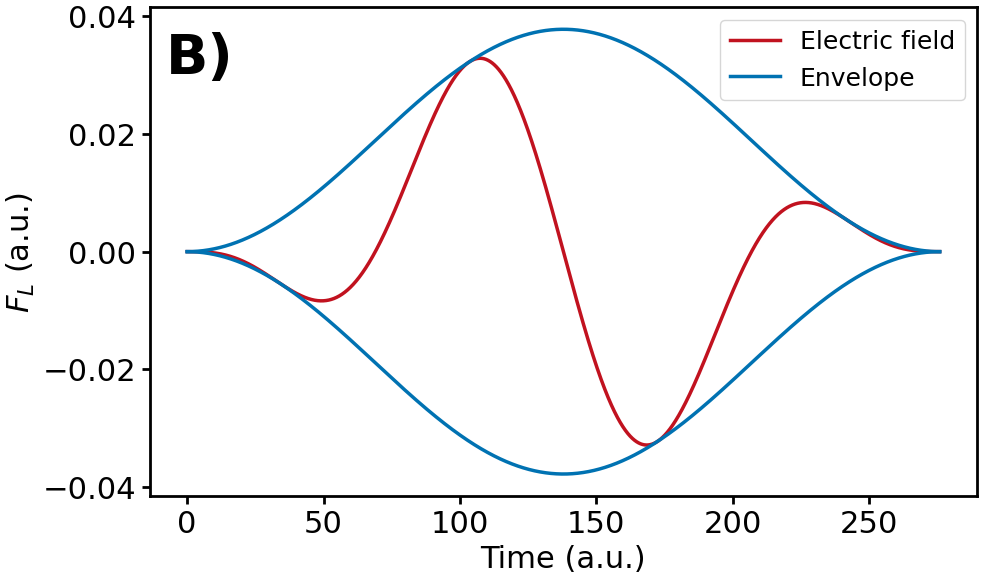}
    \caption{A) Trapezoidal and B) Sine square electric fields and their envelopes as a function of time.}
    \label{fig:efield}
\end{figure}

Two different electric field profiles are used in this article. The first is the trapezoidal laser pulse shown in Figure \ref{fig:efield} A), which takes the form:

\begin{equation}
F_L(t) = F_0 f(t) \sin(\omega_\text{L} t + \phi),
\end{equation}
where $F_{\text{0}}$ is the peak electric field strength and $\omega_\text{L}$ is the angular frequency of the laser. A phase of $\phi = \pi$ is always cosnsidered. The envelope $f(t)$ is given by:
\begin{equation}
f(t) =
\begin{cases}\dfrac{t}{t_1} & \text{if } 0 \le t < t_1 \\
1& \text{if } t_1 \le t < t_2 \\
\dfrac{\tau - t}{\tau - t_2} & \text{if } t_2 \le t \le \tau.
\end{cases}
\end{equation}
Here, the critical times of the pulse are defined in terms of the optical period $T$ as $t_1 = N_{\text{in}}T$ and $t_2 = t_1 + N_{\text{plat}}T$, with a total pulse duration of $\tau = t_2 + N_{\text{fin}}T$. The variables $N_{\text{in}}$, $N_{\text{plat}}$, and $N_{\text{fin}}$ correspond to the number of optical cycles in the initial slope, plateau, and final slope, respectively. Throughout this work, this trapezoidal pulse consists of an 8-cycle plateau and 1-cycle slopes. Unless stated otherwise, all calculations are performed at a central wavelength of 1000 nm and a peak intensity of $5.0 \times 10^{13} \text{W/cm}^2$. The cutoff's intensity and wavelength independence is explored at the end of Section \ref{sec:1cap} with the trapezoidal pulse.

The second profile is a 2-cycle sine square laser pulse, shown in Figure \ref{fig:efield} B) of the form:

\begin{equation}
F_L(t) = F_0 \sin^2\left(\frac{\pi}{\tau} t\right) \sin(\omega_\text{L} t + \phi) 
\label{eq:efield_sin2}
\end{equation}
with $\phi = \pi$, as in the trapezoidal pulse. Calculations employing the sine square field are only performed at a central wavelength of 1000 nm and a peak intensity of $5.0 \times 10^{13} \text{W/cm}^2$.

\subsection{Numerical Solution of the TDSE}
\label{sec:TDSE}

The one-dimensional code employed in this work numerically solves the TDSE. The Hamiltonian of the system, expressed in the length gauge and atomic units under the dipole and semiclassical approximations, is given by:

\begin{equation}
\hat{H}(t) = -\frac{1}{2}\frac{\partial^2}{\partial x^2} + V(x) + x F_L(t) + V_{\text{CAP}}(x) ,
\label{eq:Hamiltonian}
\end{equation}
where the first term is the kinetic energy operator, $V(x)$ is the field-free atomic or molecular potential, and $x F_L(t)$ represents the laser-matter interaction, which is time-dependent.  $V_{\text{CAP}}(x)$ is the complex absorbing potential term that is presented in Section \ref{sec:CAP}. Numerically, the wave function is represented on a spatial grid with a constant step size $\Delta x = 0.01$ a.u. The Laplacian in the kinetic energy term is evaluated using a second-order finite difference approximation, ensuring a spatial truncation error of $\mathcal{O}(\Delta x^2)$.

The time evolution is performed by advancing the system in small time steps ($\Delta t \approx 0.035$ a.u.) using the Crank-Nicolson (CN) algorithm \cite{journalarticle:Crank.Nicolson47}. This unitary approach leads to the following linear equation:

\begin{multline}
    \left[ \mathds{1} + i\hat{H}\!\left(t+ \tfrac{\Delta t}{2}\right)\tfrac{\Delta t}{2}\right]\ket{\psi(t+\Delta t)} = \\
    \left[ \mathds{1} - i\hat{H}\!\left(t+ \tfrac{\Delta t}{2}\right)\tfrac{\Delta t}{2}\right]\ket{\psi(t)}.
\end{multline}

Because the second-order spatial discretization yields a tridiagonal Hamiltonian matrix, the resulting linear system at each time step is solved efficiently using the LAPACK library \cite{book:Anderson.etal99}.

\subsection{Complex Absorbing Potential}
\label{sec:CAP}

A Complex Absorbing Potential (CAP) is an artificial imaginary term added to the Hamiltonian (\ref{eq:Hamiltonian}) in quantum numerical simulations to absorb outgoing wave packets before they reach the edges of a finite spatial grid \cite{journalarticle:Riss.Meyer93,journalarticle:Riss.Meyer95}. In the context of HHG, the strong laser field can ionize the system, generating highly energetic free electrons that travel far from the parent ion. Because our numerical model employs a finite spatial box, these wave packets could eventually reach the boundaries and reflect inward, generating unphysical interferences to the calculated signal. To prevent this, a CAP denoted as CAP$_{\text{lim}}$, taking the form \cite{,journalarticle:Santra06}:

\begin{equation}
V_{\text{CAP(lim)}}(x) = 
\begin{cases} 
-i\eta_{\text{lim}} (|x| - x_{\text{lim}})^2 \\
\hspace{2em} \text{if } |x| > x_{\text{lim}}, \\ 
0 \hspace{1em} \text{if } |x| \le x_{\text{lim}}
\end{cases}
\end{equation}

where $\eta_{\text{lim}}$ represents the strength of CAP$_{\text{lim}}$, and $x_{\text{lim}}$ denotes the boundary between CAP$_{\text{lim}}$ and the unperturbed region. Throughout this article, this  potential is included in all presented calculations. 

In some cases, an additional CAP is also introduced, located between the origin and the edge of the simulation box (i.e., preceding CAP$_{\text{lim}}$). This second potential is referred to as the ``intermediate CAP'' (CAP$_{\text{int}}$) and is only utilized when explicitly specified. Unlike the boundary CAP, this intermediate potential is placed along the electron's propagation path, meaning wave packets can approach it from either direction. To prevent unphysical spatial reflections, a symmetric profile is implemented. The intermediate CAP can be expressed as:

\begin{equation}
V_{\text{CAP(int)}}(x) =
\begin{cases}
-i\eta_{\text{int}} \left(|x| - x_{\text{int}}\right)^2 \\
\hspace{2em} \text{if } x_{\text{int}} < |x| < x_{\text{int}} + W, \\
-i\eta_{\text{int}} \left(x_{\text{int}} + 2W - |x|\right)^2 \\
\hspace{2em} \text{if } x_{\text{int}} + W \leq |x| < x_{\text{int}} + 2W, \\
0 \hspace{1em} \text{otherwise}
\end{cases}
\end{equation}

where $\eta_{\text{int}}$ is the strength of CAP$_{\text{int}}$, $x_{\text{int}}$ is the onset position where the absorption begins relative to the origin, and $W$ is half the width of the absorbing window.

\subsection{Determination of the Effective CAP Position}
\label{sec:CAP2}

There is an intrinsic uncertainty in defining the exact point where an electron is effectively absorbed, as the wave packet reaches the CAP with a broad distribution of kinetic energies. This makes the absorption depth inherently energy-dependent. However, to meaningfully compare our quantum simulations with Mondal’s spatial-limit model, a precise effective boundary position must be established. Therefore, the goal is to determine a spatial value that corresponds to where the electron is, on average, absorbed by the CAP.

To determine this effective position, the ``half-time method'' is employed for the trapezoidal pulse. This method identifies the inflection points of the norm decay corresponding to the second and third major absorption events. These absorptions are driven by the electric field during the second half of the first optical cycle (from $t_1= 69.0 \text{ a.u.}$ to $t_2= 138.0 \text{ a.u.}$ for the trapezoidal pulse in Fig. \ref{fig:efield}) and the first half of the second cycle (from $t_2= 138.0 \text{ a.u.}$ to $t_3= 207.0 \text{ a.u.}$). However, because the CAPs are located at a considerable distance from the origin, the wave packet requires time to propagate. As a result, the actual moments of maximum absorption (the ``half-times') occur with a temporal delay and often fall outside these specific laser-cycle intervals. These two specific events are selected because the initial dynamics during the very first half-cycle are still transient, while at later times, the wave-packet trajectories become increasingly overlapping and complex.

Once these two ``half-times'' are identified, the spatial probability density is calculated at both instants. The effective spatial location for each instant is found by computing the average of the probability density strictly within the CAP region. Finally, the assigned position of the CAP is defined as the average of these two spatial values. In the following, any mention of the CAP position refers to this averaged value.

\subsection{Spectral and Time-Frequency Analysis}
The HHG spectrum is obtained from the time-dependent dipole acceleration $D_\text{a}$, which is chosen instead of the dipole moment $D$ due to its reduced sensitivity to numerical noise. The acceleration is calculated from the dipole moment using Ehrenfest theorem \cite{journalarticle:Burnett.etal92} as:
\begin{equation}
    D_\text{a} (t) = \ddot{D} (t) = - \bra{\Psi (t)} \frac{\partial V}{\partial x} \ket{\Psi (t)} - F_\text{L}(t),
\end{equation}
The resulting signal is multiplied by a Hann window ($w(t)$) \cite{journalarticle:Harris78} , 
\begin{equation}
    w(t) = \frac{1}{2}\left[1 - \cos\!\left(\frac{2\pi t}{T}\right)\right],
\end{equation}
to suppress the spectral leakage caused by the finite duration of the simulation and is also zero-padded at both ends. The harmonic spectrum $S_{\mathrm{a}}(\omega)$ is then given by:

\begin{equation}
    S_{\mathrm{a}}(\omega) = \left| \int \ddot{D} (t) w(t) e^{-i\omega t} \, \dd{t} \right|^2.
\end{equation}
$S_{\mathrm{a}}(\omega)$ is evaluated numerically using a Fast Fourier Transform alogirthm, obtaining the final HHG spectra that are presented in the following section.

For the Time-Frequency analysis, the dipole acceleration is analysed using a Gabor transform, i.e. a short-time Fourier transform employing a Gaussian window \cite{journalarticle:Chirila.etal10}:

\begin{equation}
    G(t,\omega) = \int \ddot{D} (t') e^{-\frac{(t'-t)^2}{2\sigma_t^2}} e^{-i\omega t'} \, \dd{t'},
\end{equation}
where $\sigma_t$ represents the temporal width of the Gaussian window, set to 0.065 $T_\text{L}$ and $T_\text{L}$ being the optical period. The spectrogram is defined as $S(t,\omega) = |G(t,\omega)|^2$. The Gaussian window is truncated at $\pm 25\sigma_t$ , a point at which the function is effectively zero. This chosen value establishes a trade-off between temporal and spectral resolution, which is inherent to the Gabor transform. Finally, to avoid edge artifacts, the signal is multiplied by a cosine at both edges and zero-padded before performing the transform.

\section{RESULTS AND DISCUSSION}

\subsection{High-Harmonic Generation from an Isolated Water Molecule}
\label{sec:water_molecule}

\begin{figure}
    \centering
    \includegraphics[width=\linewidth]{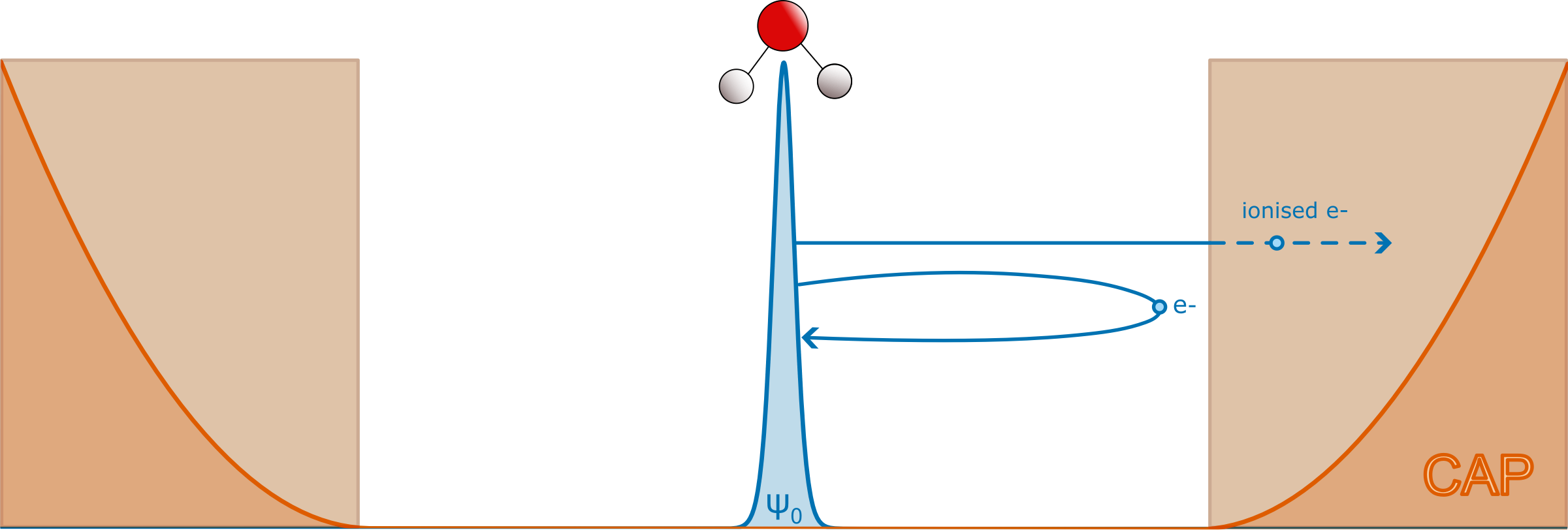}
    \caption{Schematic view of the 1D quantum model. The water molecule is centered at the origin, shown as an initial electron probability density $\Psi_0$. The arrows represent different electron trajectories in the continuum. The orange region denotes the CAP$_{\text{lim}}$ positioned at the edge of the simulation box to absorb non-returning electron trajectories (ionised electrons).}
    \label{fig:water_molec_setup}
\end{figure}

First, an isolated water molecule is studied in the gas phase. The molecular potential is centered in the simulation box, and the CAP$_{\text{lim}}$ is placed at 162.2 a.u. to ensure the complete absorption of any ionized electrons not participating in HHG. The setup is presented in Figure \ref{fig:water_molec_setup}. This CAP is set well beyond the limit of $2x_{\alpha} \approx 36.4$ a.u. Here, the parameter $x_{\alpha} = F_0/\omega_L^2$ represents a critical spatial boundary for the electron dynamics within the classical picture of HHG. As will be discussed in detail later, this boundary separates the two main families of quantum paths: short trajectories that never travel beyond $x_{\alpha}$, and long trajectories that always cross this threshold. In the classical three-step model \cite{journalarticle:Corkum93},  $2x_{\alpha}$ corresponds to the maximum distance a returning electron can travel before recombining with the parent ion. In reality, the ionized wavepacket spreads quantum mechanically and electron density can extend beyond this classical limit. However, placing the boundary at 162.2 a.u. guarantees that the HHG process remains essentially unaffected by any spatial confinement. 

\begin{figure}
    \centering
    \includegraphics[width=\linewidth]{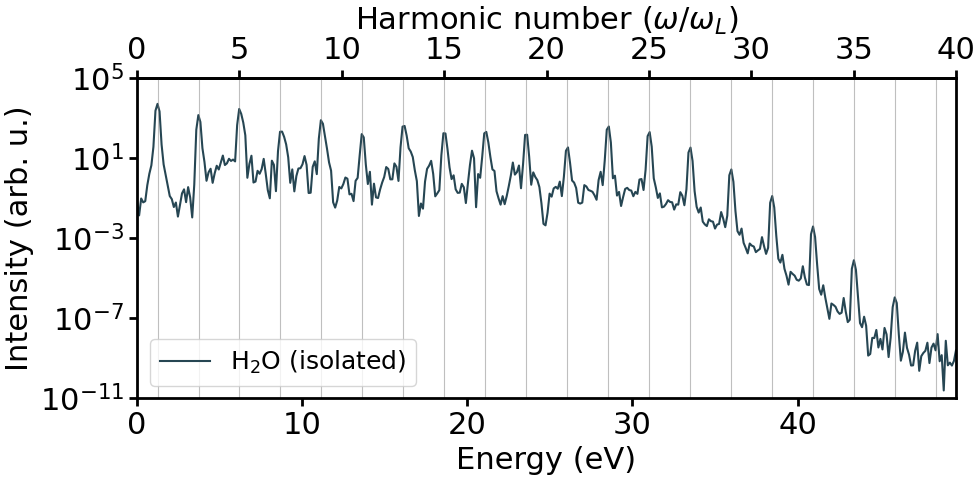}
    \caption{HHG spectrum emmitted by a 1D water molecule defined by a soft-Coulomb potential. The trapezoidal laser pulse is used with a central wavelength of 1000 nm and a peak intensity of $5.0 \times 10^{13}$ W/cm$^2$. The x-axis represents the harmonic order (the emitted frequency divided by the laser's fundamental frequency) as well as the energy (eV), and the y-axis shows the intensity in arbitrary units and logarithmic scale.}
    \label{fig:water_molec_HHG}
\end{figure}

Generated using the trapezoidal electric field, the resulting HHG spectrum is shown in Figure \ref{fig:water_molec_HHG}, where the x-axis represents the harmonic order (the emitted frequency divided by the laser's fundamental frequency) as well as the energy (eV), and the y-axis shows the intensity in arbitrary units and in logarithmic scale. The figure illustrates the typical HHG spectrum, featuring an initial slight intensity decay for the first harmonics, a plateau extending to approximately harmonic order 23 (the cutoff), and a final sharp decay. Additionally, only odd harmonics are visible, which is a direct consequence of the system's inversion symmetry. Regarding the cutoff, the semiclassical model \cite{journalarticle:Schafer.etal93, journalarticle:Corkum93} predicts it at order 22 ($\approx 21.53$), but it typically underestimates the actual position. This discrepancy partly arises because the three‑step model imposes a strict point‑like recombination condition ($x=0$). This assumption ignores the spatial spreading of the electronic wave packet \cite{journalarticle:Lewenstein.etal94}, which allows for longer trajectories and higher recombination energies.

\begin{figure}
    \centering
    \includegraphics[width=\linewidth]{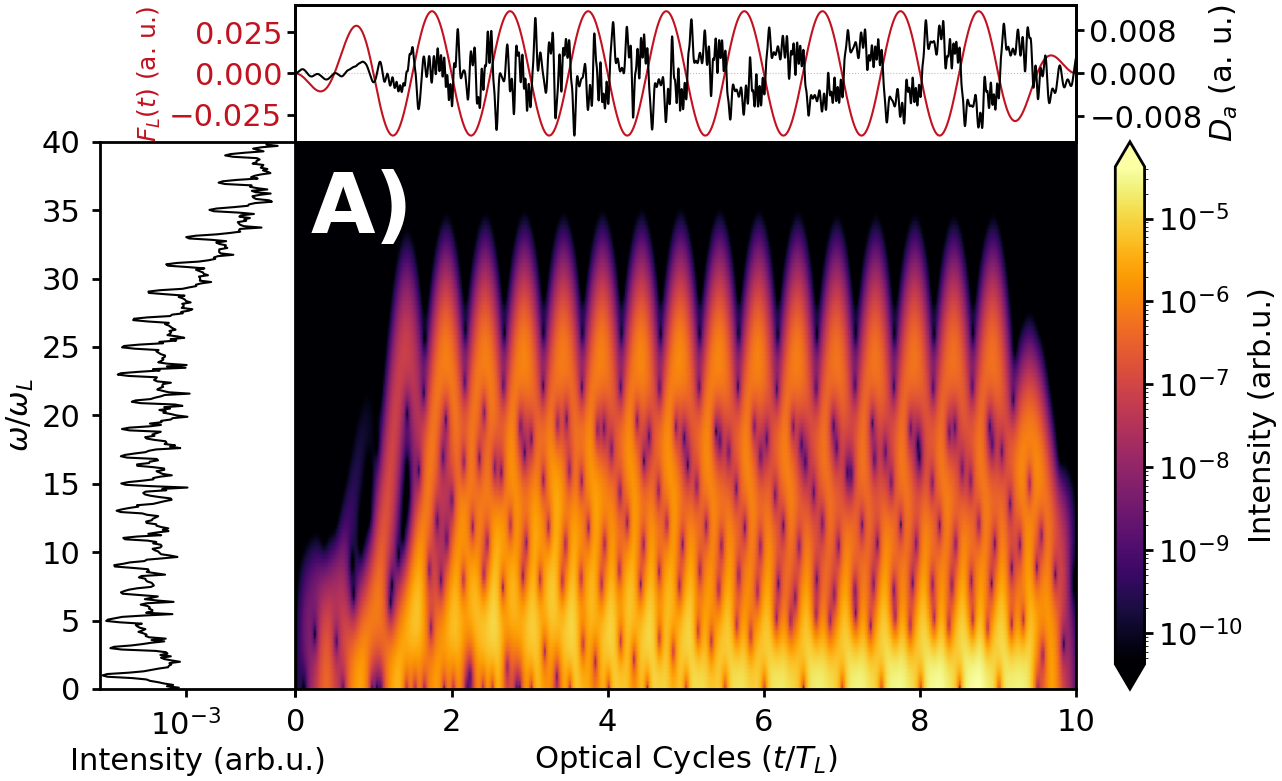}
    \includegraphics[width=\linewidth]{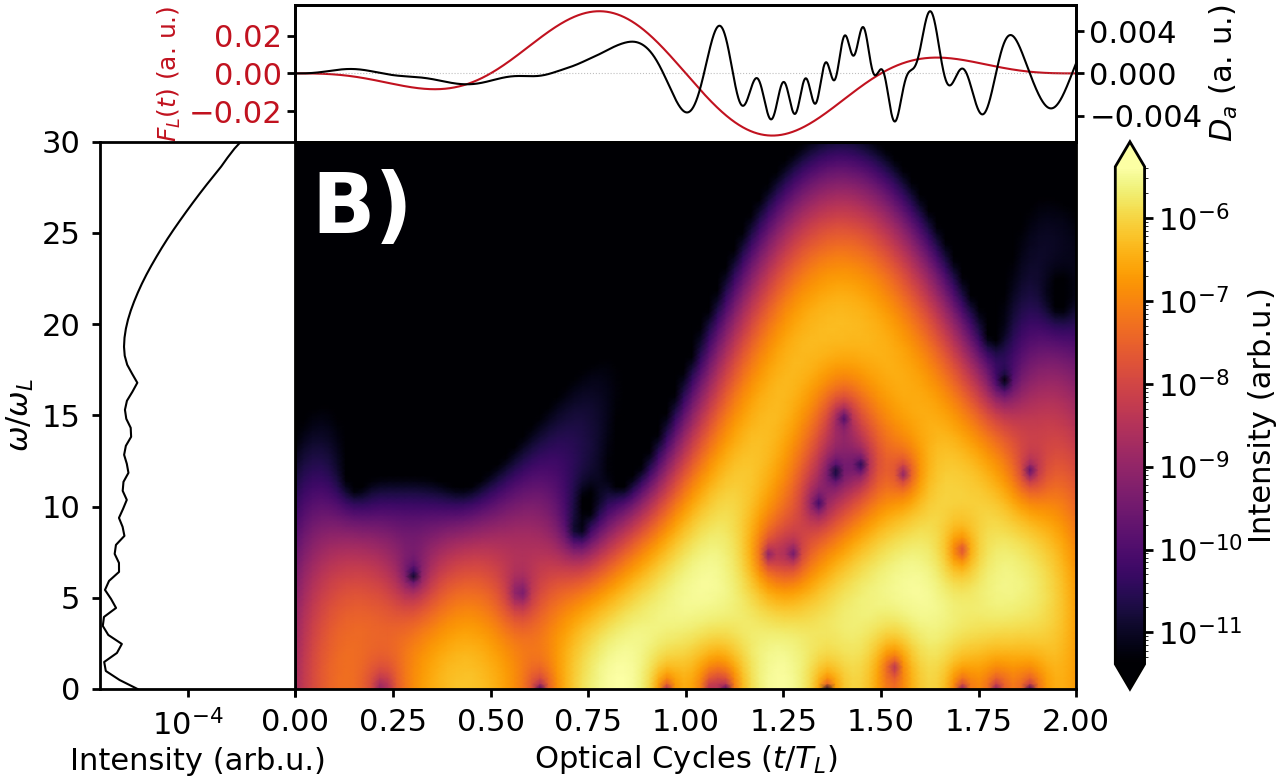} 
    \caption{Gabor transforms of the acceleration dipole $D_a$ of a water molecule under A) trapezoidal and B) sine-squared laser pulses as a function of time and frequency respectively normalized by the laser period $T_L$ and pulsation $\omega_L$. The left panel shows the HHG spectrum, with harmonic order on the y-axis and intensity on the x-axis. The top panel illustrates the driving electric field (red) and the dipole moment (black). The central panel reveals the emission times of the harmonics across the entire pulse.}
    \label{fig:water_molec_gabor}
\end{figure}

To investigate the underlying temporal dynamics of the emission process, a time-frequency analysis is performed using a Gabor transform. This technique determines the approximate emission time of each harmonic during the laser pulse. Figure \ref{fig:water_molec_gabor} A shows the resulting Gabor transform for the isolated water molecule in the trapezoidal laser pulse. The left panel displays the HHG spectrum (the same as the one shown in Figure \ref{fig:water_molec_HHG}, with harmonic order on the y-axis and intensity on the x-axis, again in logarithmic scale. The top panel illustrates the driving electric field $F_L$ (red) and the acceleration dipole moment $D_a$ (black). The central panel reveals the emission times of the harmonics across the entire trapezoidal pulse. As shown here, the emission occurs as a train of bursts appearing roughly twice per optical cycle, consistent with the three-step recollision mechanism taking place near each field extremum. Notably, the acceleration dipole response remains negligible until $t/T_L \approx 1.5$, even though the field is already well into its rising edge. This reflects the fact that efficient tunneling ionization only becomes significant once the trapezoidal envelope approaches its plateau intensity. Then, from $t/T_L \approx 1.5$ onwards, $D_a$ fluctuates very fast, until approximately $t/T_L \approx 5.0$ where the start of a periodic pattern is observed, dictated by the laser field. In the central pannel, each individual burst shows a clear chirp, as the frequency of the emitted light shifts throughout the emission time, reaching its peak at the extrema of both $F_L$ and $D_a$. Those peaks aligns with the cutoff in the HHG spectrum. This behavior is the signature of the short and long electron trajectories predicted by the semiclassical model, which ionise at slightly different moments of the field and consequently recombine with different kinetic energies. Finally, the broadened appearance of the emission is a natural consequence of the Gabor transform's time–frequency trade-off: temporal resolution was prioritized to isolate individual attosecond bursts at the expense of spectral sharpness.

To avoid the complexity of overlapping bursts and achieve a clearer time–frequency analysis, the trapezoidal pulse is replaced by a two-cycle sinusoidal one. All subsequent Gabor transform analyses are performed using this simplified configuration.
As shown in Figure \ref{fig:water_molec_gabor} B, restricting the interaction to just two optical cycles makes the short and long trajectories clearly distinguishable. In the corresponding HHG spectrum, the typical discrete odd harmonics disappear, but the cutoff remains distinctly visible. This occurs because the sine-squared pulse has a shorter temporal profile compared to the trapezoidal pulse, resulting in a single main recombination event. Furthermore, while the dipole acceleration exhibits these isolated emission bursts, the characteristic interferences between the short and long trajectories are captured in the HHG spectrum. Finally, the central panel reveals that the short trajectories exhibit a positive chirp, meaning lower frequencies are emitted first, peaking at approximately $1.40 T_L$. It is at this instant that the cutoff harmonics are generated. In contrast, the long trajectories display a negative chirp: they begin at the maximum frequency (i.e., the cutoff), coinciding with the peak of the short trajectories, and emit lower frequencies later in the pulse.

With the emission profile and characteristic electron trajectories and cutoff of the isolated water molecule already established, this gas-phase configuration provides a reliable baseline for comparison. Building upon this foundation, the following section explores how these dynamics are altered by the presence of spatial constraints. To do so, the identical laser-molecule potential setup is maintained while bringing the CAP$_{\text{lim}}$ closer to the parent ion, introducing a physical barrier into the model. This approach enables us to directly investigate how the truncation of longer trajectories, mimicking the structural confinement of a liquid environment, impacts the maximum recombination energy and the overall HHG cutoff position.

\subsection{Effect of Spatial Confinement on the HHG Cutoff}
\label{sec:1cap}

In this section, the basic setup is maintained as the one presented in Figure \ref{fig:water_molec_setup}, but this time the CAP position is varied to mimic specific aspects of the liquid-water environment, such as solvation shells. This enables the investigation of how these structural features influence the HHG cutoff position. Physically, this approach is equivalent to moving an absorbing wall that truncates any part of the electronic wave packet traveling beyond it. Consequently, as the CAP is brought closer to the molecule, electron trajectories that would normally return to the parent ion and participate in the HHG emission are absorbed. Four key distances relative to the water molecule (set at the origin) are analyzed: 36.3 a.u. and 18.3 a.u. (corresponding approximately to $2x_\alpha$ and $x_\alpha$, respectively), an intermediate distance of 10.1 a.u., and finally, 5.5 a.u. This shortest distance corresponds approximately to the oxygen-oxygen (O–O) separation between two water molecules \cite{journalarticle:Ceriotti.etal16}, representing the first solvation shell.

\begin{figure*}
    \centering
    \includegraphics[width=0.49\textwidth]{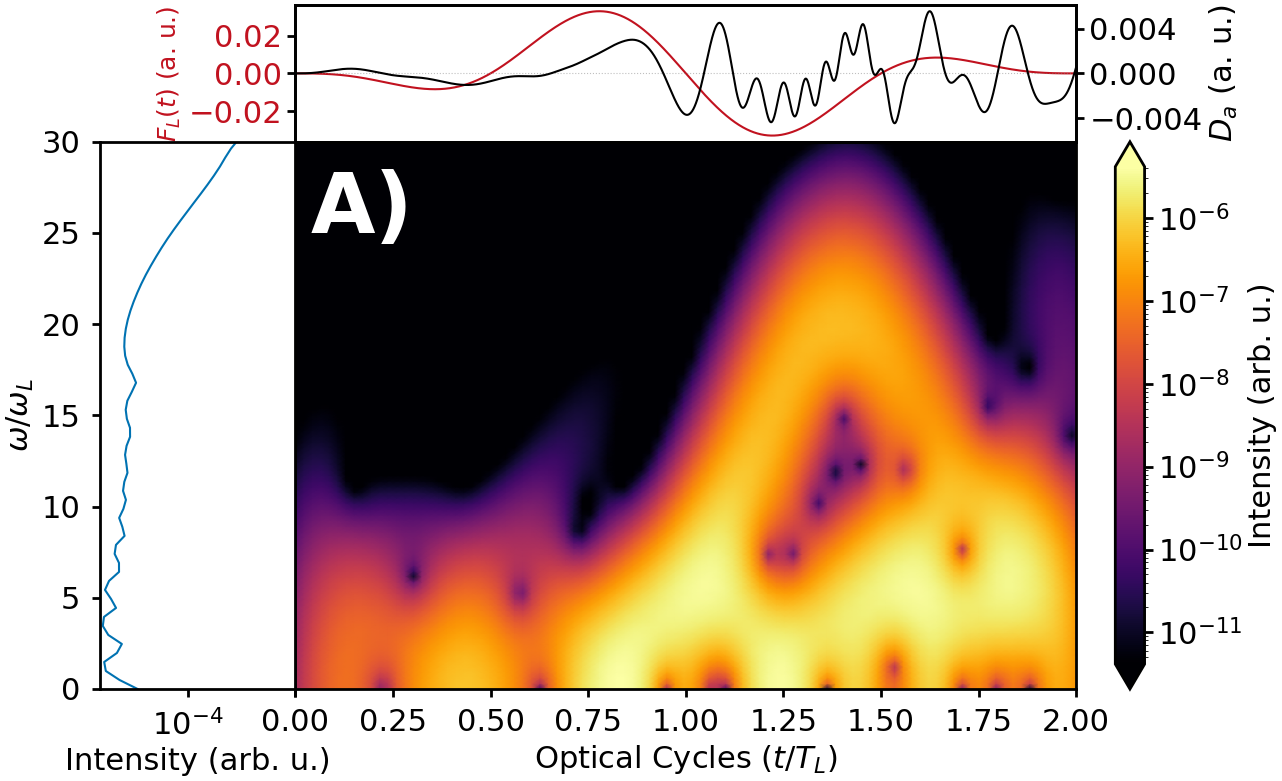}
    \includegraphics[width=0.49\textwidth]{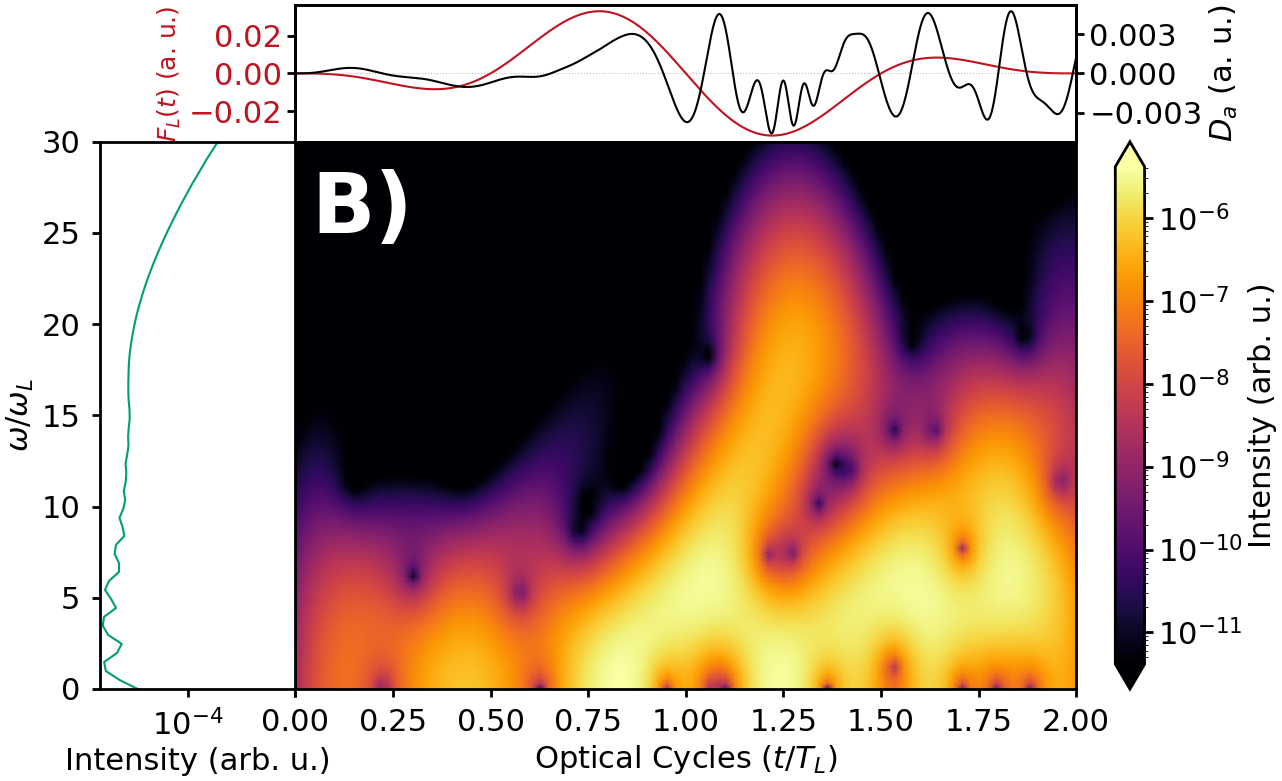}
    \includegraphics[width=0.49\textwidth]{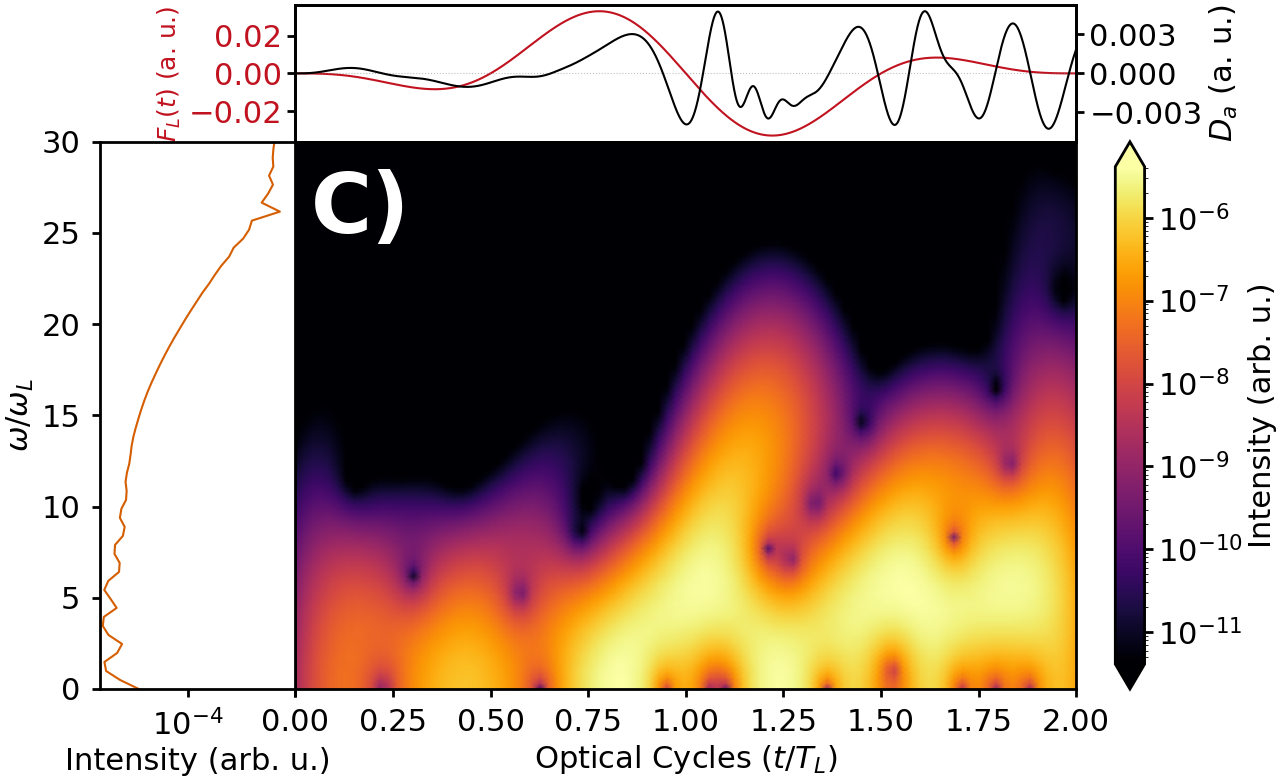}
    \includegraphics[width=0.49\textwidth]{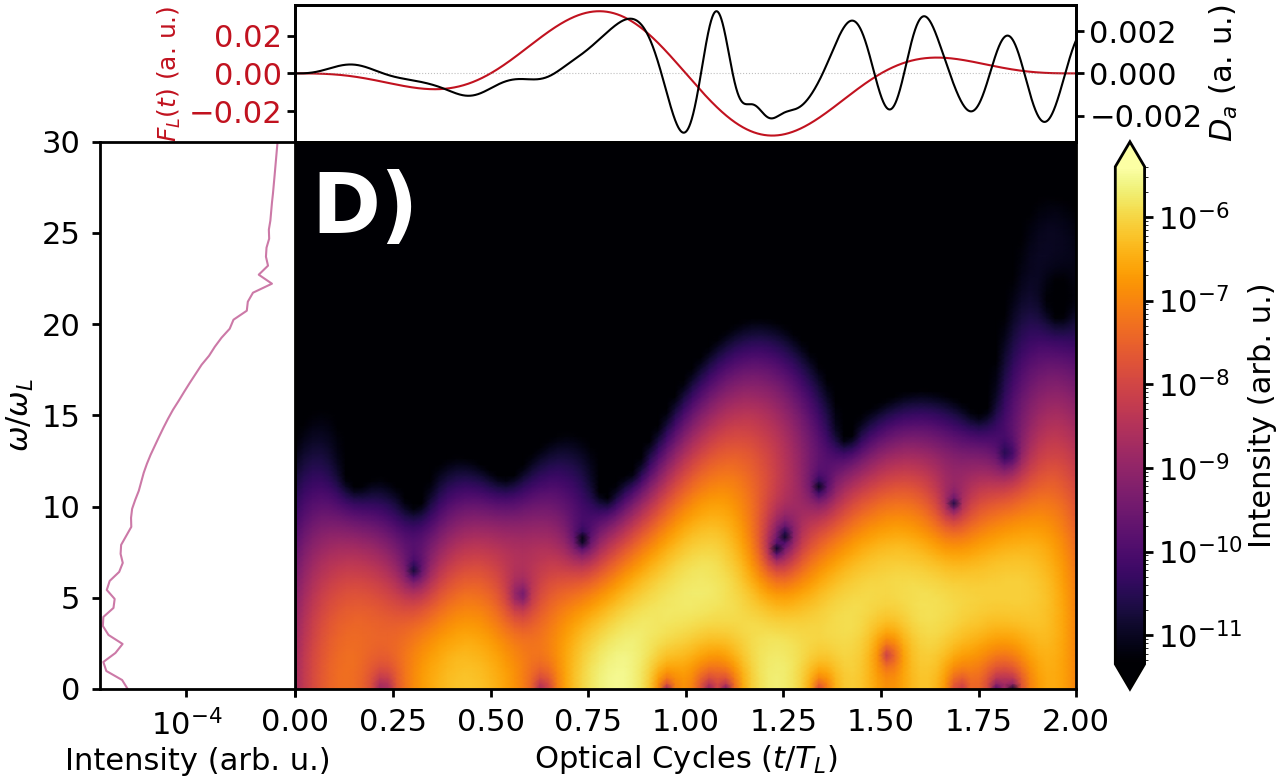}
    \caption{Gabor transforms of the acceleration dipole $D_a$ of a water molecule under the sine-squared laser pulse as a function of time and frequency respectively normalized by the laser period $T_L$ and pulsation $\omega_L$. Each panel corresponds to a different CAP$_{\text{lim}}$ position: A) 36.3 a.u., B) 18.3 a.u., C) 10.1 a.u., and D) 5.5 a.u.}
    \label{fig:Gabor_sin2_1_cap_HHG}
\end{figure*}

Figure \ref{fig:Gabor_sin2_1_cap_HHG} illustrates how the emission times evolve as a function of the CAP position for the sinusoidal pulse. In panel A, with the CAP placed at 36.3 a.u., the time-frequency structure, the acceleration dipole and the cutoff position are identical to those shown in Figure \ref{fig:water_molec_gabor} B. This is expected, as the CAP is located approximately at the limit position 2$x_{\alpha}$, allowing both long and short trajectories to fully develop. In panel B, where the CAP is placed at 18.3 a.u., the long trajectory leg of the central panel is suppressed. The dipole acceleration shows noticeably different oscillations and the HHG spectrum no longer exhibits the characteristic interference pattern between short and long trajectories. This is expected since only the short trajectories contribute to the spectrum, removing any interference with the long trajectories. However, as previously discussed in the context of Figure \ref{fig:water_molec_gabor} B, the highest-energy photons are still emitted via short trajectories. Consequently, the cutoff position remains unaffected, as it is shown in the HHG spectrum in the left panel. Finally, panels C and D reveal that as the CAP is placed closer to the water molecule, the trajectories are truncated earlier, thereby reducing the cutoff energy. The oscillations of $D_a$ decrease drastically and in the HHG spectra the decrease of the cutoff is evident. This demonstrates that the cutoff position depends directly on the location of the CAP for distances lower than $x_\alpha$. 

\begin{figure}
    \centering
    \includegraphics[width=\linewidth]{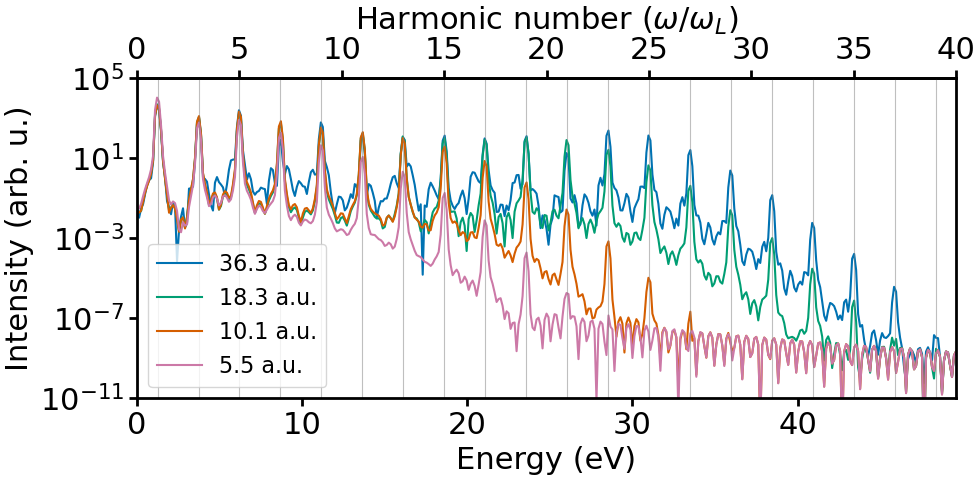}
    \caption{HHG spectra of the water molecule under the trapezoidal laser pulse for different CAP$_{\text{lim}}$ positions: 36.3 a.u. (blue), 18.3 a.u. (green), 10.1 a.u. (orange), and 5.5 a.u. (pink).} 
    \label{fig:1_cap_HHG}
\end{figure}

Figure \ref{fig:1_cap_HHG} displays the HHG spectra for these four configurations under the trapezoidal pulse. The blue and green curves, corresponding to the $2x_\alpha$ and $x_\alpha$ cases respectively, exhibit an identical cutoff at approximately harmonic order 23 ($\approx$ 29 eV). This matches the cutoff of the isolated water molecule, as anticipated from the preceding time–frequency analysis. The primary difference between these two spectra is the reduced overall intensity in the $x_\alpha$ case. This decrease is expected, as placing the CAP at $x_\alpha$ suppresses the long trajectories, therefore removing their contribution to the total emission. In the intermediate case (orange curve), the cutoff visibly decreases compared to the $x_\alpha$ configuration. The semiclassical model predicts a cutoff near harmonic 17 ($\approx$ 21 eV), which approximately aligns with the simulated cutoff observed around harmonic 15 ($\approx$ 19 eV). Finally, the pink curve corresponds to the case where the CAP is placed at approximately the first solvation shell distance. In this configuration, although the cutoff is not sharply defined, it is clearly lower than that of the intermediate case and is estimated to be around the 11th harmonic by the semiclassical model ($\approx$ 14 eV). 
Experimental results by Mondal et al. \cite{journalarticle:Mondal.etal23} estimated the first cutoff at approximately 14.2 eV (considering wavelength independence). When evaluating these results, it is crucial to distinguish between three different cutoff values: the experimental measurement (14.2 eV), the prediction of the semiclassical model ($\approx$ 14 eV), and the cutoff extracted from our 1D TDSE spectrum ($\approx$ 10 eV). 
The quantitative discrepancy between our TDSE spectrum and the experiment is  expected. Our setup isolates a single molecule with a CAP rather than modeling a macroscopic liquid, therefore neglecting complex many-body interactions and laser propagation effects. Beyond this, a rigorous quantitative comparison is complicated by several methodological factors. First, the cutoff is not defined identically across the different approaches: in our work, it is identified as the last harmonic before the intensity decay begins, whereas Mondal \textit{et al.} employ a different criterion (see Supplementary Information in \cite{journalarticle:Mondal.etal23}), making strict numerical comparisons limited. Secondly, the exact position of the CAP is also not precisely known (see section \ref{sec:CAP2}). While the semiclassical model offers a sharp, well-defined criterion for distinguishing between not only short and long trajectories but also between the returning from the non-returning trajectories, no such clean boundary exists in our quantum mechanical framework. Thirdly, high-kinetic-energy electrons are more difficult to absorb and may also partially reflect off the CAP. Although care has been taken to minimise this effect, it cannot be completely eliminated. Consequently, only qualitative agreement with the experiment is anticipated.
However, the discrepancy between our TDSE results and the semiclassical prediction is more revealing. Our TDSE calculation provides an exact quantum mechanical solution for this specific 1D potential, which means that the failure of the semiclassical model to reproduce the TDSE cutoff suggests an inherent limitation in the semiclassical approach itself. In particular, the mismatch between the semiclassically computed confinement and our quantum mechanical counterpart is significant. The discrepancy between the semi-classical picture and our TDSE results confirms once more the intrinsic quantum nature of the HHG process. While the semiclassical model is qualitatively robust, successfully explaining the intensity and wavelength independence, it lacks the necessary precision to yield exact quantitative predictions, even for a simplified 1D system. As a side note, the small, periodic peaks appearing beyond 23 eV for the pink plot are attributed to numerical noise.

\begin{figure}
    \centering
    \includegraphics[width=\linewidth]{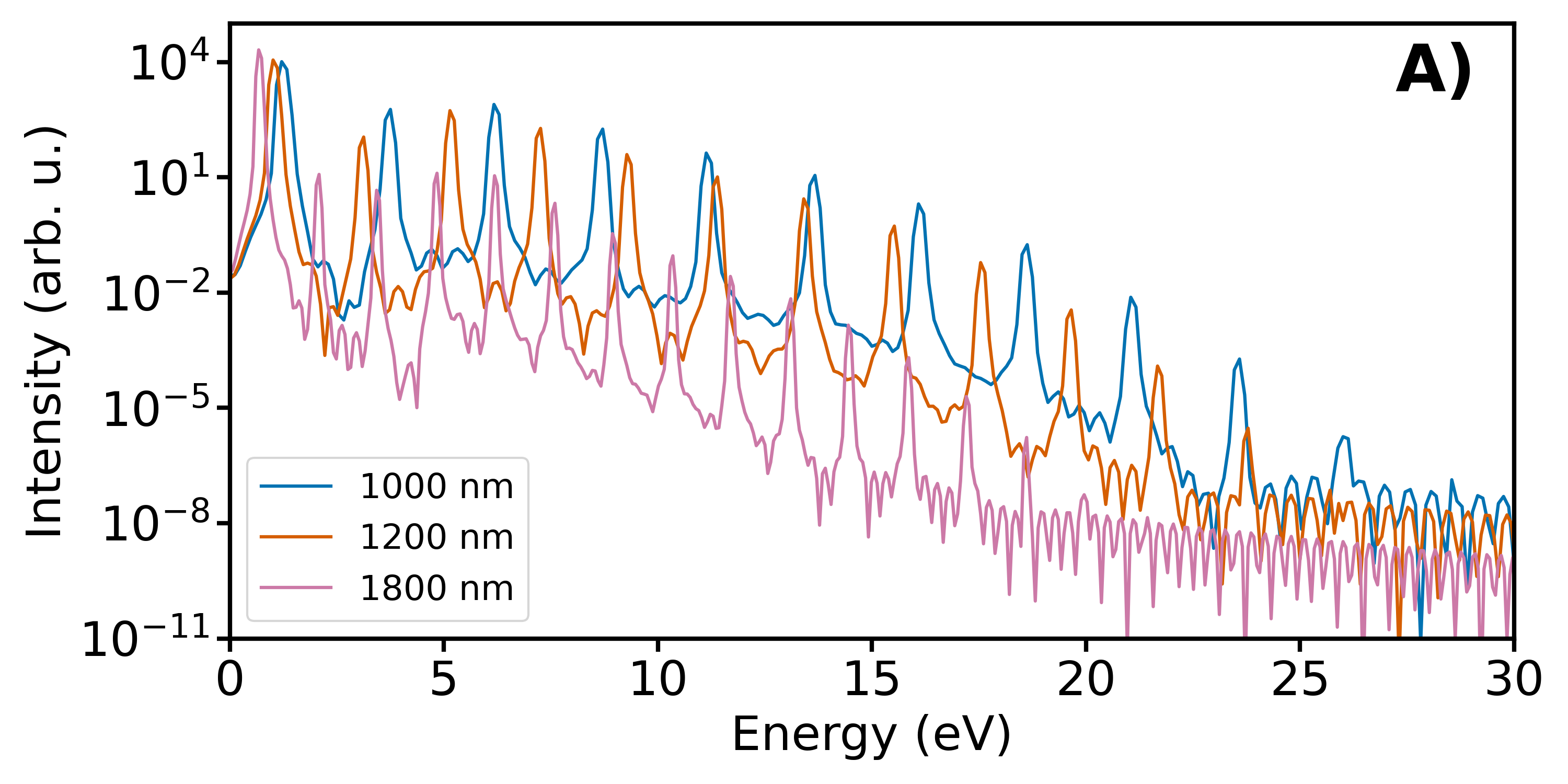}
    \includegraphics[width=\linewidth]{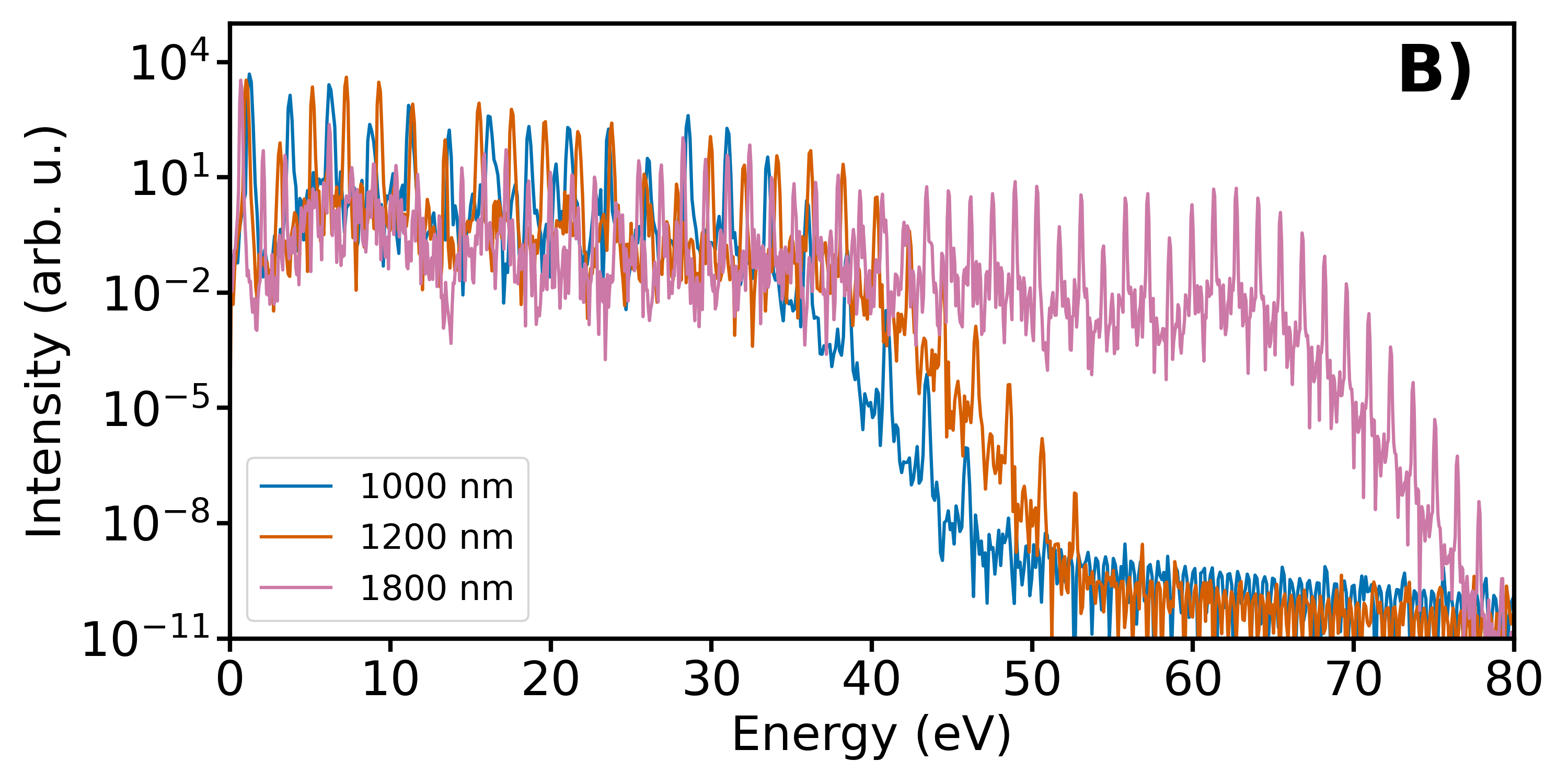}
    \caption{HHG spectra of the water molecule under a trapezoidal laser pulse with a fixed intensitity of $5.0 \times 10^{13} \text{W/cm}^2$. In panel A) CAP$_{\text{lim}}$ is placed at 5.5 a.u., representing the liquid-phase case. In panel B) CAP$_{\text{lim}}$ is placed at 162.2 a.u., representing the gas-phase case. Note that x-axis are scaled different in both panels.} 
    \label{fig:wl_HHG}
\end{figure}

\begin{figure}
    \centering
    \includegraphics[width=\linewidth]{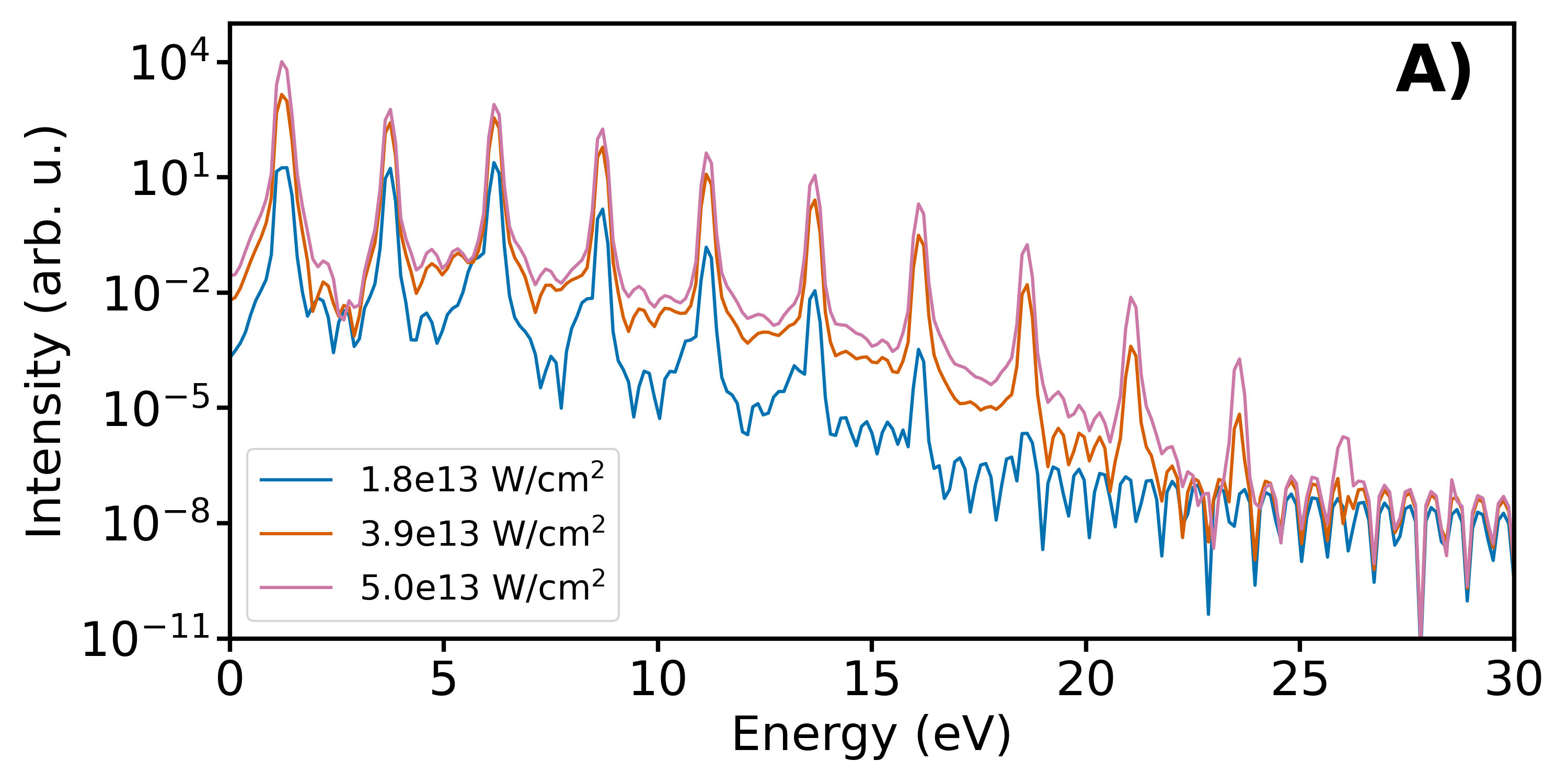}
    \includegraphics[width=\linewidth]{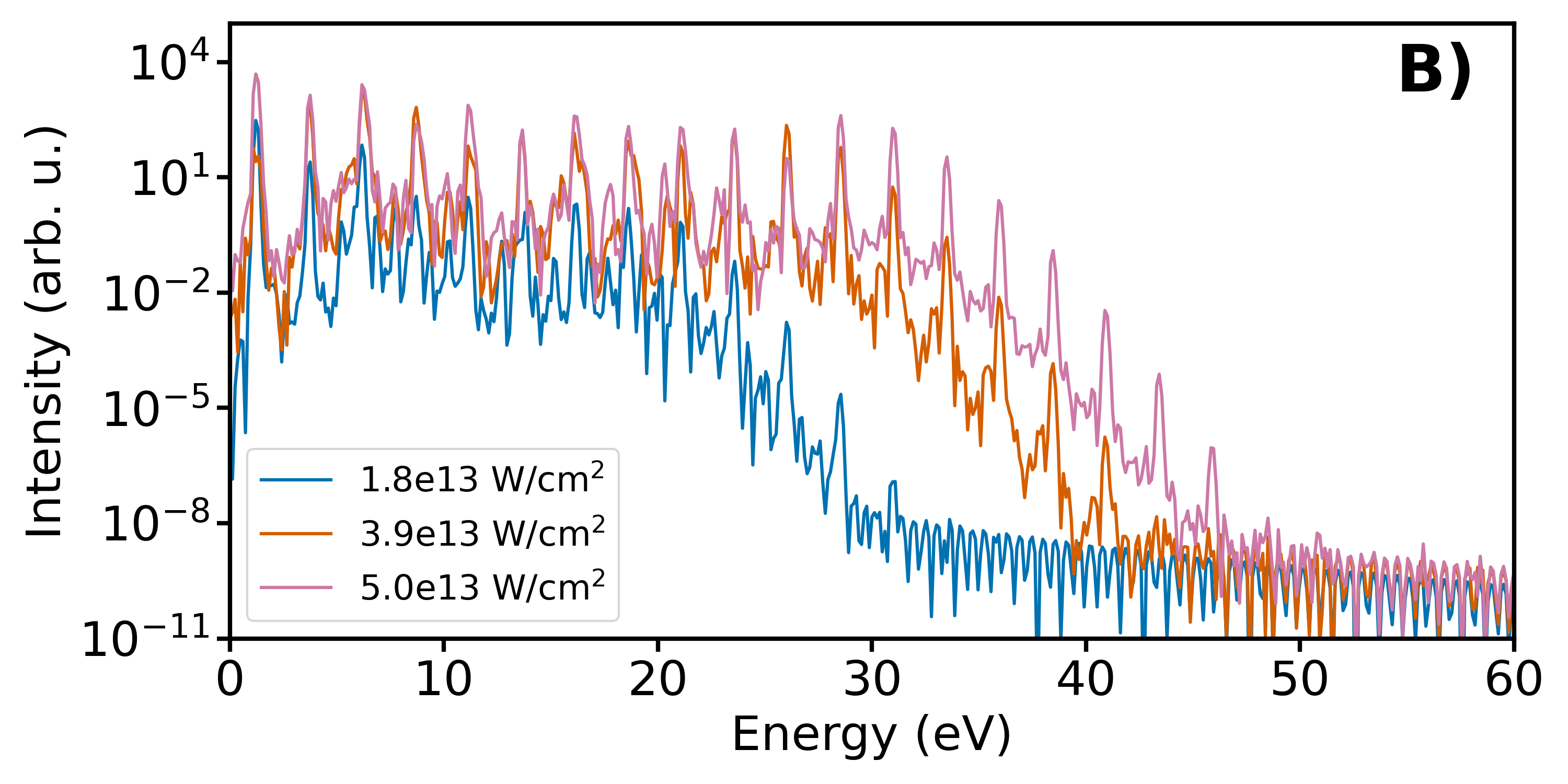} 
    \caption{HHG spectra of the water molecule under a trapezoidal laser pulse with a fixed wavelength of 1000 nm. In panel A) CAP$_{\text{lim}}$ is placed at 5.5 a.u., representing the liquid-phase case. In panel B) CAP$_{\text{lim}}$ is placed at 162.2 a.u., representing the gas-phase case. Note that x-axis are scaled different in both panels.} 
    \label{fig:int_HHG}
\end{figure}

To further validate that the physical barrier plays a dominant role in determining the maximum emission energy, the laser parameters were varied while keeping the CAP fixed at a distance of 5.5 a.u. (i.e., the first solvation shell), analogous to the studies performed by Mondal \textit{et al.} \cite{journalarticle:Mondal.etal23}. In our case, the gas-phase counterpart is also presented, with CAP$_{\text{lim}}$ placed at 162.2 a.u. In Figure \ref{fig:wl_HHG}, where the intensity is kept at $5.0 \times 10^{13} \text{W/cm}^2$ but the wavelength varies, it is shown that in the case of the liquid-phase configuration (panel A), the cutoff energy remains invariant, which is clearly not the case for the gas-phase configuration (panel B). In the latter, the cutoff energy shifts to higher values as the driving wavelength increases, which is consistent with the semiclassical model. In Figure \ref{fig:int_HHG}, a similar analysis is performed by varying the laser intensity while keeping the wavelength fixed at 1000 nm. Again, in the liquid-phase configuration (panel A), the cutoff energy remains unchanged, while in the gas-phase configuration (panel B), it shifts to higher values as the driving intensity increases. Together, these results demonstrate that the spatial restriction imposed by the CAP overrules the standard intensity and wavelength dependencies of the cutoff, firmly establishing the physical barrier as a primary limiting factor in the case of liquid-phase systems.

In summary, implementing a variable-position CAP in the 1D model provides a fully quantum framework to specifically isolate and investigate the pure effects of spatial confinement on HHG emission. As discussed in the introduction, Mondal \textit{et al.} previously proposed a semiclassical model suggesting that confinement dictates the cutoff, which aligns well with their experimental observations. Our purely quantum simulations corroborate and expand upon this idea. By systematically restricting the longer trajectories of the electronic wave packet, we demonstrate that a physical barrier directly limits the maximum recombination energy. Importantly, this approach allows us to test the exact effect of spatial truncation itself, which is fundamentally different from the complex, many-body interactions of actual neighboring molecules. Therefore, this geometric restriction provides a robust, fully quantum validation for why the primary cutoff in liquids is significantly reduced and appears to be largely independent of both the laser's wavelength and intensity.

However, while this single-barrier approach successfully explains the primary cutoff dynamics, it treats the surrounding environment as a perfectly opaque wall. In reality, solvation shells are not impenetrable boundaries, and their structural nature allows for the partial transmission of the electronic wave packet. To capture this semi-permeable behavior and investigate the origins of the experimentally observed second plateau, the following section introduces a two-CAP configuration. By allowing a fraction of the wave packet to undergo longer excursions beyond the first solvation shell, we can explore the more complex multi-plateau structure of the liquid-phase HHG spectrum.

\subsection{Simulating Water Solvation Shells: Emergence of the Second Plateau}
\label{sec:2caps}
\begin{figure}
    \centering
    \includegraphics[width=\linewidth]{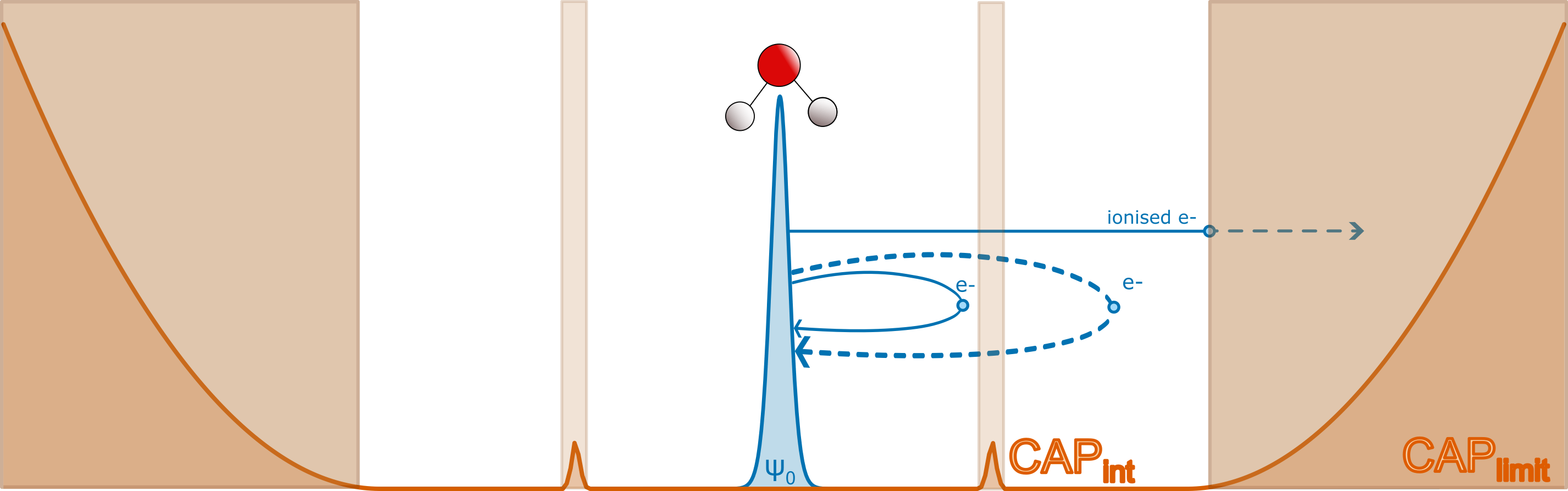}
    \caption{Same as Figure \ref{fig:water_molec_setup}, but with the addition of an intermediate CAP (CAP$_{\text{int}}$) placed between the water molecule and the fully absorbing CAP$_{\text{lim}}$. The width and strength of CAP$_{\text{int}}$ are reduced to allow partial transmission of the electronic wavefunction.}
    \label{fig:2_cap_scheme}
\end{figure}

To investigate the emergence of a second plateau, as predicted by Mondal \textit{et al.} \cite{journalarticle:Mondal.etal26}, an additional intermediate CAP (CAP$_{\text{int}}$) is introduced between the central water potential and CAP$_{\text{lim}}$. A scheme of the setup is presented in Figure \ref{fig:2_cap_scheme}, in which it is shown that the trajectories reaching a maximun distance lower than CAP$_{\text{int}}$ are kept untouched, while longer trajectories are partially absorbed by CAP$_{\text{int}}$ . The width and strength of the different CAP$_{\text{int}}$ are reduced to allow partial transmission of the electronic wavefunction while CAP$_{\text{lim}}$, at the edge of the box, is far enough so that it doesn't interfere with the HHG process, yet fully absorbs highly energetic ionised electrons.

\begin{figure}
    \centering
    \includegraphics[width=\linewidth]{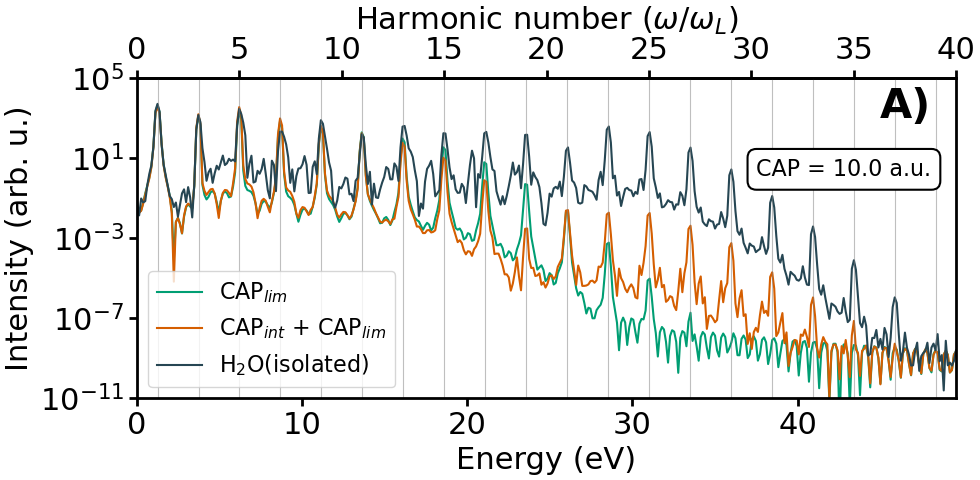}
     \includegraphics[width=\linewidth]{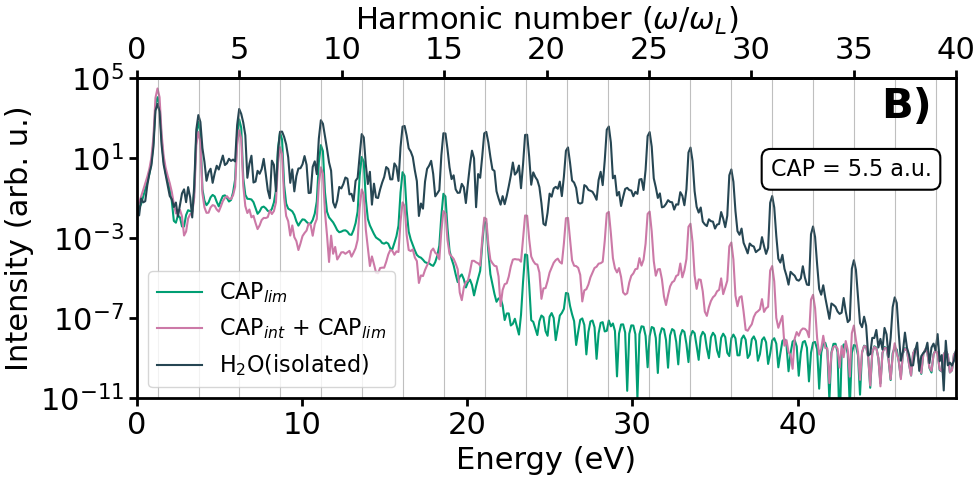}
    \caption{Comparison of the HHG spectra of the water molecule under single-CAP and two-CAP configurations. A) Spectra where the restricting barrier is placed at $10.0$ a.u. The green and black lines represent the single-CAP cases (with CAP$_{\text{lim}}$ at $10.0$ and $162.2$ a.u., respectively), while the orange line shows the two-CAP configuration, where the intermediate barrier (CAP$_{\text{int}}$) is the one located at $10.0$ a.u. B) Spectra with the restricting barrier positioned closer to the molecule at $5.5$ a.u. The green and black lines show the single-CAP (CAP$_{\text{lim}}$ at $5.5$ and $162.2$ a.u., respectively), and the pink line corresponds to the two-CAP configuration with CAP$_{\text{int}}$ at $5.5$ a.u.}
    \label{fig:2_cap_HHG}
\end{figure}

Figure \ref{fig:2_cap_HHG} displays the resulting HHG spectra when CAP$_{\text{int}}$ is positioned at 10.0 a.u. (in orange in panel A) and 5.5 a.u (in pink in panel B). In both cases, the emission clearly exhibits a two-plateau structure for the cases where the CAP$_{\text{int}}$ is used. The first plateau terminates at a primary cutoff identical to the one observed in the single-CAP configuration (in both panels, in green), governed by the spatial restriction of this first barrier. However, due to the reduced absorption of CAP$_{\text{int}}$, a secondary plateau emerges, extending to higher harmonic orders, up to around the 23rd harmonic, similar to the case of the water molecule (in both panels, in black). This distinctive double-plateau profile qualitatively reproduces the liquid-phase behavior reported by Mondal \textit{et al.} \cite{journalarticle:Mondal.etal26}.

\begin{figure*}
    \centering
    \includegraphics[width=0.49\textwidth]{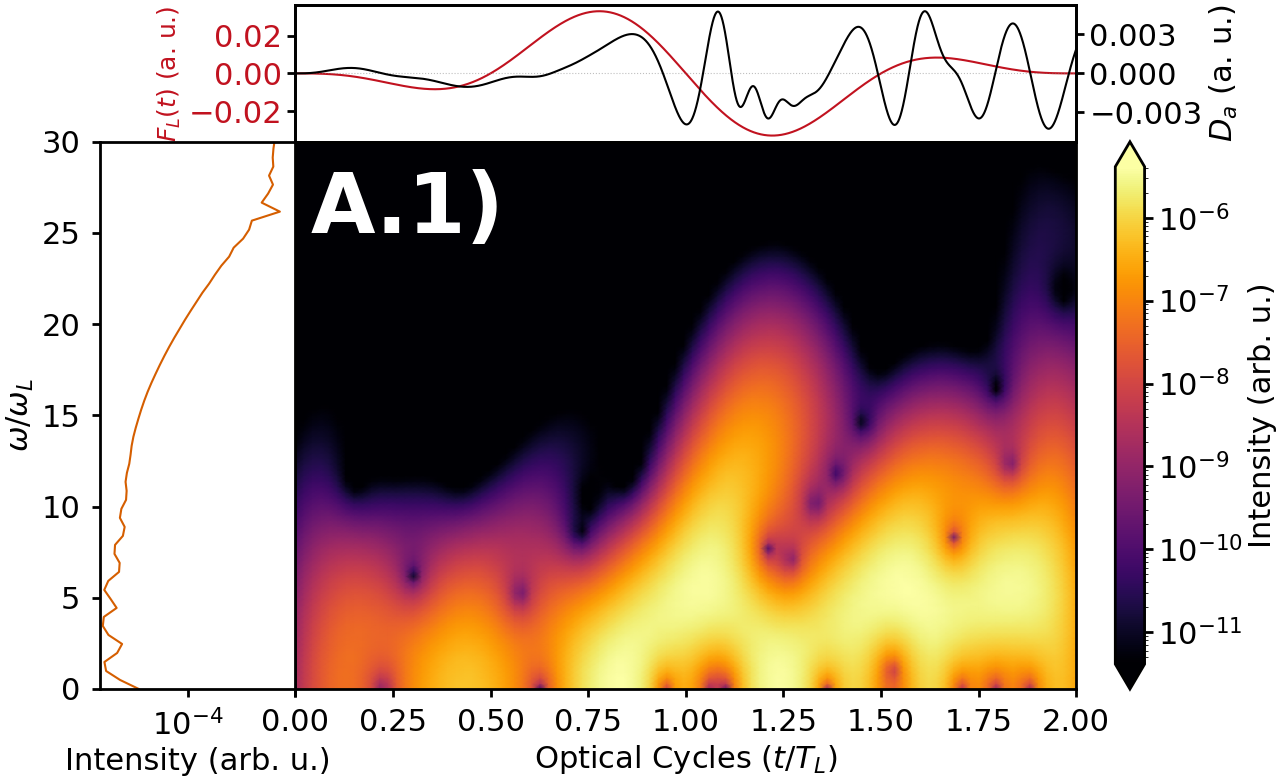}
    \includegraphics[width=0.49\textwidth]{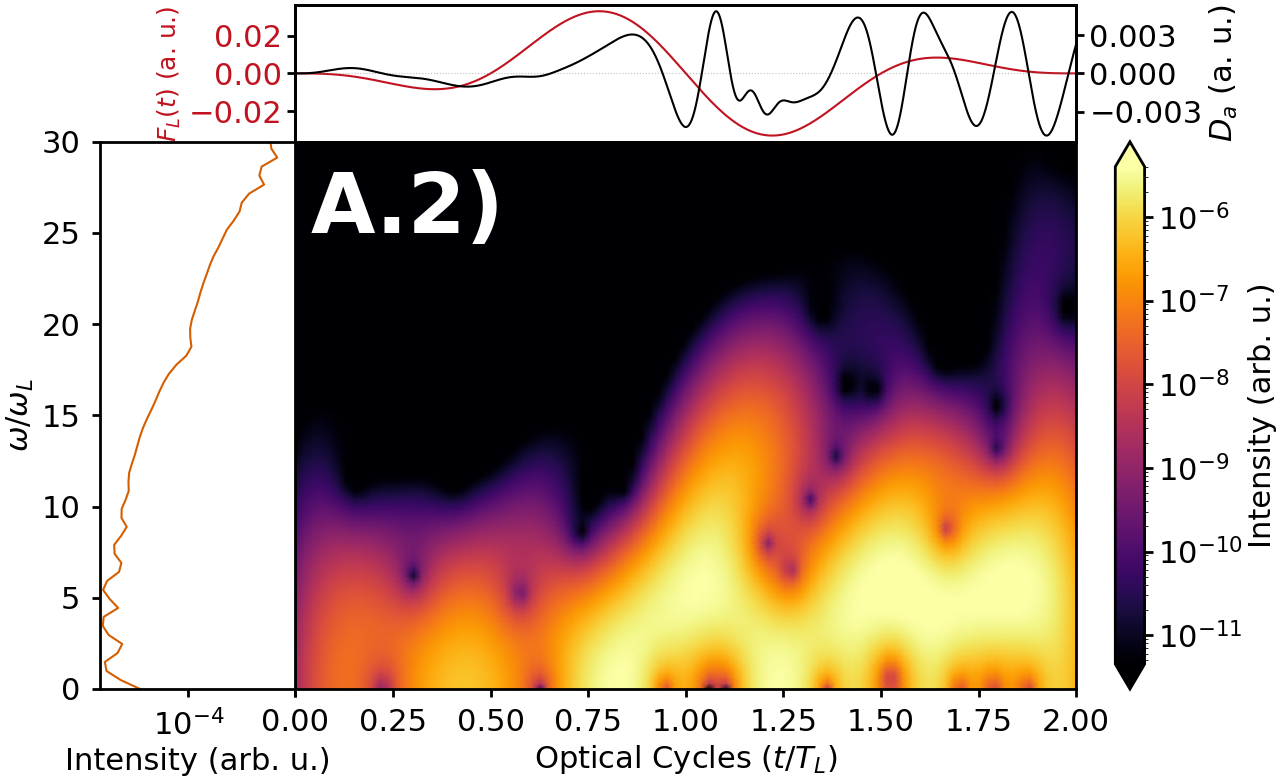}
    \includegraphics[width=0.49\textwidth]{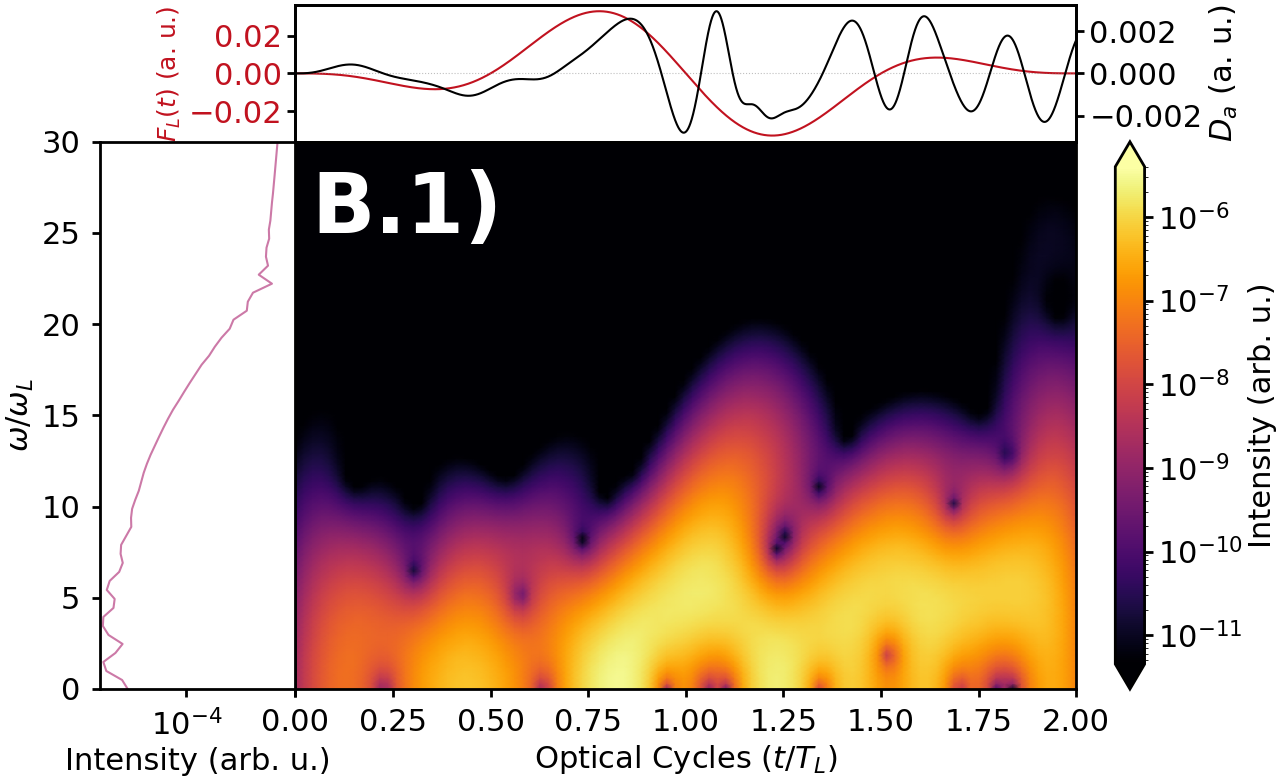}
    \includegraphics[width=0.49\textwidth]{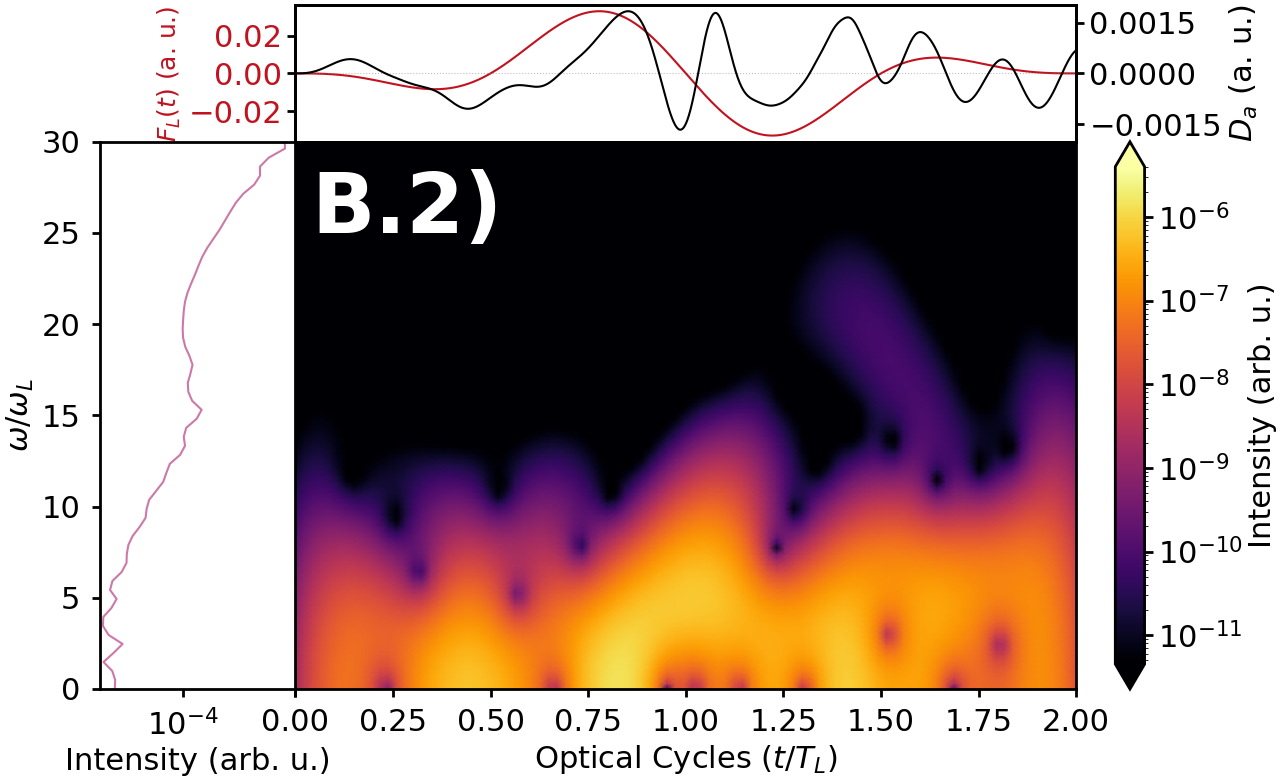}
    \caption{Gabor transforms of the acceleration dipole $D_a$ of a water molecule under the sine-squared laser pulse as a function of time and frequency respectively normalized by the laser period $T_L$ and pulsation $\omega_L$. Panel A.1) abd B.1) are taken from Figure \ref{fig:Gabor_sin2_1_cap_HHG} C) and D), respectively, and correspond to the single-CAP configuration. Panel A.2) and B.2) show the two-CAP configuration, where the intermediate barrier (CAP$_{\text{int}}$) is placed at 5.5 a.u. and 10.0 a.u., respectively.}
    \label{fig:2_cap_gabor}
\end{figure*}

Again, to understand the underlying electron dynamics responsible for this second plateau, time-frequency analysis was performed. Figure \ref{fig:2_cap_gabor} presents the corresponding Gabor transforms for the cases where CAP$_{\text{int}}$ is placed at 5.5 a.u. (panel A.2) and 10.0 a.u. (panel B.2). Panels A.1 and B.1 correspond exactly to the single-CAP results shown previously in Figures \ref{fig:2_cap_gabor} C and D, respectively, and are included here to facilitate direct comparison. In the previously discussed single-CAP scenario, long trajectories were completely suppressed by the fully absorbing barrier. In contrast, the two-CAP configuration clearly shows the emergence of the second plateau in the HHG spectra. Looking at the dipole acceleration, a change in the oscillation pattern is observed, which strongly hints at the return of extended trajectories. This is confirmed in the central panel: the time-frequency structure reveals that these long trajectories are partially restored, appearing as faint purple signals. These signals are responsible for the emission in the second plateau. 

Since CAP$_{\text{int}}$ only partially absorbs the wave packet, a fraction of the electron density is transmitted, travels further into the continuum, and successfully returns to the parent ion. These surviving fractions undergo longer excursions in the continuum and thus acquire higher kinetic energies before recombination. The faint time-frequency signature of these extended trajectories confirms that their highest-energy components are directly responsible for the emission in the second plateau. Furthermore, the reduced electron density of this transmitted wave packet explains why the overall intensity of this second plateau is significantly lower than that of the first. Physically, this demonstrates that the liquid environment acts as a series of semi-permeable structural barriers (solvation shells) rather than a single, opaque wall. Electrons that penetrate beyond the first solvation shell can reach greater distances, naturally giving rise to multiple plateaus and cutoffs in the HHG spectrum. Looking forward, this link between wave-packet transmission and emission yield opens an interesting perspective: by comparing the intensity ratio between the first and second plateaus, it may be possible to extract an estimate for the effective ``electron permeability" of the first solvation shell in liquid samples.

\section{CONCLUSION}
\label{sec:conclusion}

In this work, we have investigated the role of physical constraints on HHG in liquid water using a fully quantum-mechanical one-dimensional model. By varying the position and strength of CAP$_{\text{lim}}$, we transitioned from simulating an isolated gas-phase water molecule to emulating the spatially confined environment characteristic of a liquid. Our simulations demonstrate that spatial confinement plays a dominant role in determining the maximum emission energy of an electron in liquid-phase HHG. The physical truncation of the electron's trajectories by the nearest neighboring molecules limits the maximum recombination energy, explaining the significant cutoff reduction observed compared to the gas phase. This spatial restriction provides a consistent physical mechanism that explains why the primary HHG cutoff in liquids is largely independent of both the driving laser's intensity and wavelength, unlike the standard behavior observed in gas-phase systems. In addition, by introducing a semi-permeable CAP$_{\text{int}}$, we attempted to model the structural nature of solvation shells in 1D. This configuration allows a small fraction of the electronic wave packet to transmit through the first barrier and undergo longer excursions in the continuum. Time-frequency analysis confirmed that the highest-energy components of these surviving trajectories are  responsible for the emergence of a secondary, less intense plateau that reaches the same cutoff energy as the isolated water molecule.

Despite these findings, we must acknowledge the inherent limitations of the present study. While the one-dimensional nature of the model is a notable approximation, it is not the only factor limiting exact quantitative predictions. In this initial work, our primary objective was to isolate the pure effect of spatial truncation. As a result, we have not yet attempted to estimate the impact of structural disorder or potential multi-electron effects, both of which are intrinsically present in a real liquid phase. Our method is threfore designed to capture the fundamental qualitative behavior of the system rather than provide exact quantitative predictions. To move beyond these constraints, future efforts will be directed toward investigating these disorder and many-body phenomena, along with fully three-dimensional simulations that, although computationally more demanding, will allow for a more accurate representation of the liquid environment and its influence on HHG. 

\section*{DATA AVAILABILITY}

The data supporting the findings of this study are openly available on Zenodo at \url{https://doi.org/10.5281/zenodo.23390572}.

\clearpage
\bibliographystyle{apsrev}
\bibliography{article1d}

@book{book:Anderson.etal99,
  title = {{{LAPACK Users}}' {{Guide}}},
  author = {Anderson, E. and Bai, Z. and Bischof, C. and Blackford, L. S. and Demmel, J. and Dongarra, J. and Du Croz, J. and Greenbaum, A. and Hammarling, S. and McKenney, A. and Sorensen, D.},
  year = 1999,
  month = jan,
  edition = {Third},
  publisher = {{Society for Industrial and Applied Mathematics}},
  doi = {10.1137/1.9780898719604},
  urldate = {2026-09-21},
  langid = {english}
}

@article{journalarticle:Awasthi.etal08,
  title = {Single-Active-Electron Approximation for Describing Molecules in Ultrashort Laser Pulses and Its Application to Molecular Hydrogen},
  author = {Awasthi, Manohar and Vanne, Yulian V. and Saenz, Alejandro and Castro, Alberto and Decleva, Piero},
  year = 2008,
  month = jun,
  journal = {Phys. Rev. A},
  volume = {77},
  number = {6},
  pages = {063403},
  publisher = {American Physical Society},
  doi = {10.1103/PhysRevA.77.063403},
  urldate = {2026-09-30}
}

@article{journalarticle:Ceriotti.etal16,
  title = {Nuclear {{Quantum Effects}} in {{Water}} and {{Aqueous Systems}}: {{Experiment}}, {{Theory}}, and {{Current Challenges}}},
  shorttitle = {Nuclear {{Quantum Effects}} in {{Water}} and {{Aqueous Systems}}},
  author = {Ceriotti, Michele and Fang, Wei and Kusalik, Peter G. and McKenzie, Ross H. and Michaelides, Angelos and Morales, Miguel A. and Markland, Thomas E.},
  year = 2016,
  month = jul,
  journal = {Chem. Rev.},
  volume = {116},
  number = {13},
  pages = {7529--7550},
  doi = {10.1021/acs.chemrev.5b00674},
  urldate = {2026-01-21},
  langid = {english}
}

@article{journalarticle:Chen.Bian23,
  title = {Theoretical Analysis of High-Order Harmonic Generation in Liquids by a Semiclassical Method},
  author = {Chen, Jia-Xiang and Bian, Xue-Bin},
  year = 2023,
  month = apr,
  journal = {Phys. Rev. A},
  volume = {107},
  number = {4},
  pages = {043111},
  doi = {10.1103/PhysRevA.107.043111},
  urldate = {2026-03-31},
  langid = {english}
}

@article{journalarticle:Corkum.Krausz07,
  title = {Attosecond Science},
  author = {Corkum, P. B. and Krausz, Ferenc},
  year = 2007,
  month = jun,
  journal = {Nat. Phys.},
  volume = {3},
  number = {6},
  pages = {381--387},
  publisher = {Nature Publishing Group},
  doi = {10.1038/nphys620},
  urldate = {2026-09-14},
  copyright = {2007 Springer Nature Limited},
  langid = {english}
}

@article{journalarticle:Corkum93,
  title = {Plasma Perspective on Strong Field Multiphoton Ionization},
  author = {Corkum, P. B.},
  year = 1993,
  month = sep,
  journal = {Phys. Rev. Lett.},
  volume = {71},
  number = {13},
  pages = {1994--1997},
  doi = {10.1103/PhysRevLett.71.1994},
  urldate = {2026-04-08},
  copyright = {http://link.aps.org/licenses/aps-default-license},
  langid = {english}
}

@article{journalarticle:Crank.Nicolson47,
  title = {A Practical Method for Numerical Evaluation of Solutions of Partial Differential Equations of the Heat-Conduction Type},
  author = {Crank, J. and Nicolson, P.},
  year = 1947,
  month = jan,
  journal = {Math. Proc. Camb. Philos. Soc.},
  volume = {43},
  number = {1},
  pages = {50--67},
  doi = {10.1017/S0305004100023197},
  urldate = {2026-06-30},
  langid = {english}
}

@article{journalarticle:Ferray.etal88,
  title = {Multiple-Harmonic Conversion of 1064 Nm Radiation in Rare Gases},
  author = {Ferray, M. and L'Huillier, A. and Li, X. F. and Lompre, L. A. and Mainfray, G. and Manus, C.},
  year = 1988,
  month = feb,
  journal = {J. Phys. B At. Mol. Opt. Phys.},
  volume = {21},
  number = {3},
  pages = {L31},
  doi = {10.1088/0953-4075/21/3/001},
  urldate = {2026-09-14},
  langid = {english}
}

@article{journalarticle:Gaumnitz.etal17,
  title = {Streaking of 43-Attosecond Soft-{{X-ray}} Pulses Generated by a Passively {{CEP-stable}} Mid-Infrared Driver},
  author = {Gaumnitz, Thomas and Jain, Arohi and Pertot, Yoann and Huppert, Martin and Jordan, Inga and {Ardana-Lamas}, Fernando and W{\"o}rner, Hans Jakob},
  year = 2017,
  month = oct,
  journal = {Opt. Express},
  volume = {25},
  number = {22},
  pages = {27506--27518},
  publisher = {Optica Publishing Group},
  doi = {10.1364/OE.25.027506},
  urldate = {2026-09-14},
  copyright = {\&\#169; 2017 Optical Society of America},
  langid = {english}
}

@article{journalarticle:Ghimire.etal11,
  title = {Observation of High-Order Harmonic Generation in a Bulk Crystal},
  author = {Ghimire, Shambhu and DiChiara, Anthony D. and Sistrunk, Emily and Agostini, Pierre and DiMauro, Louis F. and Reis, David A.},
  year = 2011,
  month = feb,
  journal = {Nat. Phys.},
  volume = {7},
  number = {2},
  pages = {138--141},
  publisher = {Nature Publishing Group},
  doi = {10.1038/nphys1847},
  urldate = {2026-09-14},
  copyright = {2010 Springer Nature Limited},
  langid = {english}
}

@article{journalarticle:Ghimire.Reis19a,
  title = {High-Harmonic Generation from Solids},
  author = {Ghimire, Shambhu and Reis, David A.},
  year = 2019,
  month = jan,
  journal = {Nat. Phys.},
  volume = {15},
  number = {1},
  pages = {10--16},
  doi = {10.1038/s41567-018-0315-5},
  urldate = {2025-01-31},
  langid = {english}
}

@article{journalarticle:Goulielmakis.Brabec22,
  title = {High Harmonic Generation in Condensed Matter},
  author = {Goulielmakis, Eleftherios and Brabec, Thomas},
  year = 2022,
  month = jun,
  journal = {Nat. Photonics},
  volume = {16},
  number = {6},
  pages = {411--421},
  publisher = {Nature Publishing Group},
  doi = {10.1038/s41566-022-00988-y},
  urldate = {2026-09-14},
  copyright = {2022 Springer Nature Limited},
  langid = {english}
}

@article{journalarticle:Guan.etal16,
  title = {High-Order-Harmonic Generation from Periodic Potentials Driven by Few-Cycle Laser Pulses},
  author = {Guan, Zhong and Zhou, Xiao-Xin and Bian, Xue-Bin},
  year = 2016,
  month = mar,
  journal = {Phys. Rev. A},
  volume = {93},
  number = {3},
  pages = {033852},
  publisher = {American Physical Society},
  doi = {10.1103/PhysRevA.93.033852},
  urldate = {2026-07-01}
}

@article{journalarticle:Hentschel.etal01a,
  title = {Attosecond Metrology},
  author = {Hentschel, M. and Kienberger, R. and Spielmann, Ch and Reider, G. A. and Milosevic, N. and Brabec, T. and Corkum, P. and Heinzmann, U. and Drescher, M. and Krausz, F.},
  year = 2001,
  month = nov,
  journal = {Nature},
  volume = {414},
  number = {6863},
  pages = {509--513},
  publisher = {Nature Publishing Group},
  doi = {10.1038/35107000},
  urldate = {2026-09-14},
  copyright = {2001 Macmillan Magazines Ltd.},
  langid = {english}
}

@article{journalarticle:Labeye.etal18,
  title = {Optimal {{Basis Set}} for {{Electron Dynamics}} in {{Strong Laser Fields}}: {{The}} Case of {{Molecular Ion H2}}+},
  shorttitle = {Optimal {{Basis Set}} for {{Electron Dynamics}} in {{Strong Laser Fields}}},
  author = {Labeye, Marie and Zapata, Felipe and Coccia, Emanuele and V{\'e}niard, Val{\'e}rie and Toulouse, Julien and Caillat, J{\'e}r{\'e}mie and Ta{\"i}eb, Richard and Luppi, Eleonora},
  year = 2018,
  month = nov,
  journal = {J. Chem. Theory Comput.},
  volume = {14},
  number = {11},
  pages = {5846--5858},
  doi = {10.1021/acs.jctc.8b00656},
  urldate = {2026-09-30}
}

@article{journalarticle:Lewenstein.etal94,
  title = {Theory of High-Harmonic Generation by Low-Frequency Laser Fields},
  author = {Lewenstein, M. and Balcou, {\relax Ph}. and Ivanov, M. {\relax Yu}. and L'Huillier, Anne and Corkum, P. B.},
  year = 1994,
  month = mar,
  journal = {Phys. Rev. A},
  volume = {49},
  number = {3},
  pages = {2117--2132},
  doi = {10.1103/PhysRevA.49.2117},
  urldate = {2026-06-16},
  copyright = {http://link.aps.org/licenses/aps-default-license},
  langid = {english}
}

@article{journalarticle:Luu.etal15,
  title = {Extreme Ultraviolet High-Harmonic Spectroscopy of Solids},
  author = {Luu, T. T. and Garg, M. and Kruchinin, S. Yu and Moulet, A. and Hassan, M. Th and Goulielmakis, E.},
  year = 2015,
  month = may,
  journal = {Nature},
  volume = {521},
  number = {7553},
  pages = {498--502},
  publisher = {Nature Publishing Group},
  doi = {10.1038/nature14456},
  urldate = {2026-09-14},
  copyright = {2015 Springer Nature Limited},
  langid = {english}
}

@article{journalarticle:Luu.etal18,
  title = {Extreme--Ultraviolet High--Harmonic Generation in Liquids},
  author = {Luu, Tran Trung and Yin, Zhong and Jain, Arohi and Gaumnitz, Thomas and Pertot, Yoann and Ma, Jun and W{\"o}rner, Hans Jakob},
  year = 2018,
  month = sep,
  journal = {Nat. Commun.},
  volume = {9},
  number = {1},
  pages = {3723},
  publisher = {Nature Publishing Group},
  doi = {10.1038/s41467-018-06040-4},
  urldate = {2025-01-27},
  copyright = {2018 The Author(s)},
  langid = {english}
}

@article{journalarticle:McPherson.etal87,
  title = {Studies of Multiphoton Production of Vacuum-Ultraviolet Radiation in the Rare Gases},
  author = {McPherson, A. and Gibson, G. and Jara, H. and Johann, U. and Luk, T. S. and McIntyre, I. A. and Boyer, K. and Rhodes, C. K.},
  year = 1987,
  month = apr,
  journal = {J. Opt. Soc. Am. B},
  volume = {4},
  number = {4},
  pages = {595--601},
  publisher = {Optica Publishing Group},
  doi = {10.1364/JOSAB.4.000595},
  urldate = {2026-09-14},
  copyright = {\copyright{} 1987 Optical Society of America},
  langid = {english}
}

@article{journalarticle:Mondal.etal23,
  title = {High-Harmonic Spectroscopy of Low-Energy Electron-Scattering Dynamics in Liquids},
  author = {Mondal, Angana and Neufeld, Ofer and Yin, Zhong and Nourbakhsh, Zahra and Svoboda, V{\'i}t and Rubio, Angel and {Tancogne-Dejean}, Nicolas and W{\"o}rner, Hans Jakob},
  year = 2023,
  month = dec,
  journal = {Nat. Phys.},
  volume = {19},
  number = {12},
  pages = {1813--1820},
  publisher = {Nature Publishing Group},
  doi = {10.1038/s41567-023-02214-0},
  urldate = {2024-11-13},
  copyright = {2023 The Author(s)},
  langid = {english}
}

@article{journalarticle:Mondal.etal26,
  title = {Multi-Plateau High-Harmonic Generation in Liquids Driven by off-Site Recombination},
  author = {Mondal, Angana and Neufeld, Ofer and Bal{\v c}i{\=u}nas, Tadas and Waser, Benedikt and M{\"u}ller, Serge and Rossi, Mariana and Yin, Zhong and Rubio, Angel and {Tancogne-Dejean}, Nicolas and W{\"o}rner, Hans Jakob},
  year = 2026,
  month = feb,
  journal = {Nat. Photonics},
  volume = {20},
  number = {2},
  pages = {216--224},
  publisher = {Nature Publishing Group},
  doi = {10.1038/s41566-025-01805-y},
  urldate = {2026-02-18},
  copyright = {2025 The Author(s)},
  langid = {english}
}

@article{journalarticle:Neufeld.etal22,
  title = {Ab {{Initio Cluster Approach}} for {{High Harmonic Generation}} in {{Liquids}}},
  author = {Neufeld, Ofer and Nourbakhsh, Zahra and {Tancogne-Dejean}, Nicolas and Rubio, Angel},
  year = 2022,
  month = jul,
  journal = {J. Chem. Theory Comput.},
  volume = {18},
  number = {7},
  pages = {4117--4126},
  doi = {10.1021/acs.jctc.2c00235},
  urldate = {2024-11-14},
  copyright = {https://creativecommons.org/licenses/by/4.0/},
  langid = {english}
}

@article{journalarticle:Osika.etal17,
  title = {Wannier-{{Bloch Approach}} to {{Localization}} in {{High-Harmonics Generation}} in {{Solids}}},
  author = {Osika, Edyta N. and Chac{\'o}n, Alexis and Ortmann, Lisa and Su{\'a}rez, Noslen and {P{\'e}rez-Hern{\'a}ndez}, Jose Antonio and Szafran, Bart{\l}omiej and Ciappina, Marcelo F. and Sols, Fernando and Landsman, Alexandra S. and Lewenstein, Maciej},
  year = 2017,
  month = may,
  journal = {Phys. Rev. X},
  volume = {7},
  number = {2},
  pages = {021017},
  publisher = {American Physical Society},
  doi = {10.1103/PhysRevX.7.021017},
  urldate = {2026-09-14}
}

@article{journalarticle:Page.etal88,
  title = {High-Resolution Photoionization Spectrum of Water Molecules in a Supersonic Beam},
  author = {Page, Ralph H. and Larkin, Robert J. and Shen, Y. R. and Lee, Y. T.},
  year = 1988,
  month = feb,
  journal = {J. Chem. Phys.},
  volume = {88},
  number = {4},
  pages = {2249--2263},
  doi = {10.1063/1.454058},
  urldate = {2025-10-01},
  langid = {english}
}

@article{journalarticle:Paul.etal01,
  title = {Observation of a {{Train}} of {{Attosecond Pulses}} from {{High Harmonic Generation}}},
  author = {Paul, P. M. and Toma, E. S. and Breger, P. and Mullot, G. and Aug{\'e}, F. and Balcou, {\relax Ph}. and Muller, H. G. and Agostini, P.},
  year = 2001,
  month = jun,
  journal = {Science},
  volume = {292},
  number = {5522},
  pages = {1689--1692},
  publisher = {American Association for the Advancement of Science},
  doi = {10.1126/science.1059413},
  urldate = {2026-09-14}
}

@article{journalarticle:Pertot.etal17,
  title = {Time-Resolved x-Ray Absorption Spectroscopy with a Water Window High-Harmonic Source},
  author = {Pertot, Yoann and Schmidt, C{\'e}dric and Matthews, Mary and Chauvet, Adrien and Huppert, Martin and Svoboda, Vit and {von Conta}, Aaron and Tehlar, Andres and Baykusheva, Denitsa and Wolf, Jean-Pierre and W{\"o}rner, Hans Jakob},
  year = 2017,
  month = jan,
  journal = {Science},
  volume = {355},
  number = {6322},
  pages = {264--267},
  publisher = {American Association for the Advancement of Science},
  doi = {10.1126/science.aah6114},
  urldate = {2026-09-14}
}

@article{journalarticle:Riss.Meyer93,
  title = {Calculation of Resonance Energies and Widths Using the Complex Absorbing Potential Method},
  author = {Riss, U. V. and Meyer, H.-D.},
  year = 1993,
  month = dec,
  journal = {J. Phys. B At. Mol. Opt. Phys.},
  volume = {26},
  number = {23},
  pages = {4503},
  doi = {10.1088/0953-4075/26/23/021},
  urldate = {2026-09-30},
  langid = {english}
}

@article{journalarticle:Riss.Meyer95,
  title = {Reflection-Free Complex Absorbing Potentials},
  author = {Riss, U. V. and Meyer, H.-D.},
  year = 1995,
  month = apr,
  journal = {J. Phys. B At. Mol. Opt. Phys.},
  volume = {28},
  number = {8},
  pages = {1475},
  doi = {10.1088/0953-4075/28/8/012},
  urldate = {2026-09-30},
  langid = {english}
}

@article{journalarticle:Santra06,
  title = {Why Complex Absorbing Potentials Work: {{A}} Discrete-Variable-Representation Perspective},
  shorttitle = {Why Complex Absorbing Potentials Work},
  author = {Santra, Robin},
  year = 2006,
  month = sep,
  journal = {Phys. Rev. A},
  volume = {74},
  number = {3},
  pages = {034701},
  doi = {10.1103/PhysRevA.74.034701},
  urldate = {2026-09-30},
  copyright = {http://link.aps.org/licenses/aps-default-license},
  langid = {english}
}

@article{journalarticle:Schafer.etal93,
  title = {Above Threshold Ionization beyond the High Harmonic Cutoff},
  author = {Schafer, K. J. and Yang, Baorui and DiMauro, L. F. and Kulander, K. C.},
  year = 1993,
  month = mar,
  journal = {Phys. Rev. Lett.},
  volume = {70},
  number = {11},
  pages = {1599--1602},
  doi = {10.1103/PhysRevLett.70.1599},
  urldate = {2026-06-16},
  copyright = {http://link.aps.org/licenses/aps-default-license},
  langid = {english}
}

@article{journalarticle:Vampa.etal15a,
  title = {Semiclassical Analysis of High Harmonic Generation in Bulk Crystals},
  author = {Vampa, G. and McDonald, C. R. and Orlando, G. and Corkum, P. B. and Brabec, T.},
  year = 2015,
  month = feb,
  journal = {Phys. Rev. B},
  volume = {91},
  number = {6},
  pages = {064302},
  publisher = {American Physical Society},
  doi = {10.1103/PhysRevB.91.064302},
  urldate = {2026-09-14}
}

@article{journalarticle:Wituschek.etal20,
  title = {Tracking Attosecond Electronic Coherences Using Phase-Manipulated Extreme Ultraviolet Pulses},
  author = {Wituschek, Andreas and Bruder, Lukas and Allaria, Enrico and Bangert, Ulrich and Binz, Marcel and Borghes, Roberto and Callegari, Carlo and Cerullo, Giulio and Cinquegrana, Paolo and Giannessi, Luca and Danailov, Miltcho and Demidovich, Alexander and Di Fraia, Michele and Drabbels, Marcel and Feifel, Raimund and Laarmann, Tim and Michiels, Rupert and Mirian, Najmeh Sadat and Mudrich, Marcel and Nikolov, Ivaylo and O'Shea, Finn H. and Penco, Giuseppe and Piseri, Paolo and Plekan, Oksana and Prince, Kevin Charles and Przystawik, Andreas and Ribi{\v c}, Primo{\v z} Rebernik and Sansone, Giuseppe and Sigalotti, Paolo and Spampinati, Simone and Spezzani, Carlo and Squibb, Richard James and Stranges, Stefano and Uhl, Daniel and Stienkemeier, Frank},
  year = 2020,
  month = feb,
  journal = {Nat. Commun.},
  volume = {11},
  number = {1},
  pages = {883},
  publisher = {Nature Publishing Group},
  doi = {10.1038/s41467-020-14721-2},
  urldate = {2026-09-14},
  copyright = {2020 The Author(s)},
  langid = {english}
}

@article{journalarticle:Wu.etal16,
  title = {Multilevel Perspective on High-Order Harmonic Generation in Solids},
  author = {Wu, Mengxi and Browne, Dana A. and Schafer, Kenneth J. and Gaarde, Mette B.},
  year = 2016,
  month = dec,
  journal = {Phys. Rev. A},
  volume = {94},
  number = {6},
  pages = {063403},
  publisher = {American Physical Society},
  doi = {10.1103/PhysRevA.94.063403},
  urldate = {2026-09-14}
}

@article{journalarticle:Yang.etal21,
  title = {Strong-Field Coherent Control of Isolated Attosecond Pulse Generation},
  author = {Yang, Yudong and Mainz, Roland E. and Rossi, Giulio Maria and Scheiba, Fabian and {Silva-Toledo}, Miguel A. and Keathley, Phillip D. and Cirmi, Giovanni and K{\"a}rtner, Franz X.},
  year = 2021,
  month = nov,
  journal = {Nat. Commun.},
  volume = {12},
  number = {1},
  pages = {6641},
  publisher = {Nature Publishing Group},
  doi = {10.1038/s41467-021-26772-0},
  urldate = {2026-09-14},
  copyright = {2021 The Author(s)},
  langid = {english}
}

@article{journalarticle:Yin.etal23,
  title = {Femtosecond Proton Transfer in Urea Solutions Probed by {{X-ray}} Spectroscopy},
  author = {Yin, Zhong and Chang, Yi-Ping and Bal{\v c}i{\=u}nas, Tadas and Shakya, Yashoj and Djorovi{\'c}, Aleksa and Gaulier, Geoffrey and Fazio, Giuseppe and Santra, Robin and Inhester, Ludger and Wolf, Jean-Pierre and W{\"o}rner, Hans Jakob},
  year = 2023,
  month = jul,
  journal = {Nature},
  volume = {619},
  number = {7971},
  pages = {749--754},
  publisher = {Nature Publishing Group},
  doi = {10.1038/s41586-023-06182-6},
  urldate = {2026-09-14},
  copyright = {2023 The Author(s)},
  langid = {english}
}

@article{journalarticle:Yue.Gaarde20,
  title = {Imperfect {{Recollisions}} in {{High-Harmonic Generation}} in {{Solids}}},
  author = {Yue, Lun and Gaarde, Mette B.},
  year = 2020,
  month = apr,
  journal = {Phys. Rev. Lett.},
  volume = {124},
  number = {15},
  pages = {153204},
  publisher = {American Physical Society},
  doi = {10.1103/PhysRevLett.124.153204},
  urldate = {2026-09-14}
}

@article{journalarticle:Zeng.Bian20,
  title = {Impact of {{Statistical Fluctuations}} on {{High Harmonic Generation}} in {{Liquids}}},
  author = {Zeng, Ai-Wu and Bian, Xue-Bin},
  year = 2020,
  month = may,
  journal = {Phys. Rev. Lett.},
  volume = {124},
  number = {20},
  pages = {203901},
  doi = {10.1103/PhysRevLett.124.203901},
  urldate = {2026-06-30},
  langid = {english}
}

@phdthesis{thesis:Labeye18,
  title = {Molecules Interacting with Short and Intense Laser Pulses : Simulations of Correlated Ultrafast Dynamics},
  author = {Labeye, Marie},
  year = 2018,
  school = {Sorbonne Universit\'e}
}

@article{journalarticle:Burnett.etal92,
  title = {Calculation of the Background Emitted during High-Harmonic Generation},
  author = {Burnett, K. and Reed, V. C. and Cooper, J. and Knight, P. L.},
  year = 1992,
  month = mar,
  journal = {Phys. Rev. A},
  volume = {45},
  number = {5},
  pages = {3347--3349},
  publisher = {American Physical Society},
  doi = {10.1103/PhysRevA.45.3347},
  urldate = {2026-10-06}
}

@article{journalarticle:Harris78,
  title = {On the Use of Windows for Harmonic Analysis with the Discrete {{Fourier}} Transform},
  author = {Harris, F.J.},
  year = 1978,
  month = jan,
  journal = {Proc. IEEE},
  volume = {66},
  number = {1},
  pages = {51--83},
  doi = {10.1109/PROC.1978.10837},
  urldate = {2026-10-06}
}

@article{journalarticle:Chirila.etal10,
  title = {Emission Times in High-Order Harmonic Generation},
  author = {Chiril{\u a}, C. C. and Dreissigacker, I. and {van der Zwan}, E. V. and Lein, M.},
  year = 2010,
  month = mar,
  journal = {Phys. Rev. A},
  volume = {81},
  number = {3},
  pages = {033412},
  publisher = {American Physical Society},
  doi = {10.1103/PhysRevA.81.033412},
  urldate = {2026-10-06}
}

@article{journalarticle:Ding.Bian26,
  title = {Role of a Hydrogen-Bond Network in Liquid High-Order Harmonic Generation},
  author = {Ding, Zhuang-Wei and Bian, Xue-Bin},
  year = 2026,
  month = oct,
  journal = {Phys. Rev. A},
  volume = {114},
  number = {4},
  pages = {043110},
  doi = {10.1103/m94b-82wb},
  urldate = {2026-10-16},
  langid = {english}
}

\end{document}